\documentstyle[12pt,aaspp4]{article}

\begin{document}

\def\kms{km~s$^{-1}$}
\def\msun{$M_{\odot}$}
\def\rsun{$R_{\odot}$}
\def\lsun{$L_{\odot}$}
\def\halpha{H$\alpha$}
\def\Teff{T$_{\rm eff}$}
\def\logg{$log_g$}

\tighten

\title {A search for L dwarf binary systems}

\author{I. Neill Reid\altaffilmark{1}, John E. Gizis\altaffilmark{2},
J. Davy Kirkpatrick\altaffilmark{2}, D. W. Koerner\altaffilmark{1}}

\altaffiltext{1}{Department of Physics and Astronomy, University of
Pennsylvania, 209 South 33rd St., Philadelphia, PA 19104-6396,
inr@morales.physics.upenn.edu}

\altaffiltext{2}{Infrared Processing and Analysis Center, MS 100-22, California
Institute of Technology, Pasadena, CA 91125, davy@ipac.caltech.edu,
gizis@ipac.caltech.edu}

\begin{abstract} We present analysis of HST Planetary Camera images of twenty L
dwarfs identified in the course of the Two Micron All-Sky Survey. Four
of the targets, 2MASSW J0746425+200032, 2MASSs J0850359+105716, 2MASSW J0920122+351742
and 2MASSW J1146345+223053, have
faint, red companions at separations between 0.07 and 0.29 arcseconds
(1.6 to 7.6 AU). Ground-based infrared imaging confirms the last as a 
common proper-motion companion.  
The surface density of background sources with comparable colours
is extremely low, and we identify all four as physical binaries. In three cases,
the bolometric magnitudes of the components differ by less than 0.3 magnitudes.
Since the cooling rate for brown dwarfs is a strong function of mass, similarity
in luminosities implies comparable masses. The faint component in the
2M0850 system, however, is over 1.3 magnitudes fainter than the primary in
the I-band, and $\sim0.8$ magnitudes fainter in M$_{bol}$. Indeed, 
2M0850B is $\sim0.8$ magnitudes fainter in I than the lowest luminosity L
dwarf currently known, while the absolute magnitude we deduce at J is 
almost identical with M$_J$ for Gl 229B. 
We discuss the implications of these results for the temperature
scale in the L/T transition region. 2M0850 is known to exhibit 
$\lambda 6708$\AA\ Li I absorption, indicating that the primary has a mass 
less than 0.06M$_\odot$.
Theoretical models predict that the magnitude difference implies
 a mass ratio of $\approx$0.75.

The apparent binary fraction of the current sample, 20\%, is comparable with
the results of previous surveys of late-type M dwarfs in the field and in the Hyades
cluster. However, the mean separation of the L dwarf binaries in the current sample
is smaller by a factor of two than the M dwarf value, and only one system would
be detected at the distance of the Hyades.
We discuss the likely binary frequency amongst L dwarfs in light of these new data.

\end{abstract}


\vfill\eject

\section {Introduction}

The hypothesis that stars might form  gravitationally-bound binary
systems dates to 1783 with John Goodricke's explanation of the periodic
luminosity variations in both $\beta$ Persei (Algol) and $\delta$ Cephei. Subsequent
observations (reported by Pickering in 1881) confirmed at least the former case, and the frequency of
binarity as a function of spectral type is now recognised as a significant constraint on
global star formation theories. Moreover, binarity may influence
the characteristics of planetary systems by affecting the structure and extent of 
protoplanetary disks. 

The frequency of stellar-mass companions varies with the mass of the primary. The
multiple star fraction (msf) is defined as the fraction of stellar systems which 
are binary or multiple.  Studies of solar-type stars (Duquennoy
\& Mayor, 1991) indicate a high msf, exceeding 60\%. In contrast, surveys
of lower-mass M dwarfs find a multiplicity frequency of 35\%, with companions
at separations from 0.01 to 2500 AU  (Fischer \& Marcy, 1992;  Reid \& Gizis, 1997a, b). 
Approximately 5\% of M dwarfs in the Solar Neighbourhood are companions of 
more massive stars (spectral  type K and earlier). Overall, 60\% of M dwarf systems are
single.

In contrast, recent observations have resulted in the
identification of a surprisingly large number of binaries amongst the lower-temperature,
lower-mass L dwarfs discovered in the course of the new generation of near-infrared sky surveys.
Two of the three L dwarfs in the original DENIS brown dwarf mini-survey (Delfosse {\sl et al.}, 1997)
prove to be binary (Mart\'in {\sl et al.}, 1999; Koerner {\sl et al.}, 1999 - hereinafter Ko99). 
Three of ten L dwarfs observed by Ko99 using the Near-Infrared Camera 
(NIRC) on Keck are identified  as binaries. 

At first sight, these results suggest a high binary fraction for L dwarf systems.
However, almost all of the binaries detected to date 
are spatially-resolved systems with equal-luminosity
components at separations of $<0.5$ arcseconds, unresolved in the original infrared
surveys. Both the DENIS (Delfosse {\sl et al.}, 1997) and 2MASS (Kirkpatrick {\sl et al.}, 1999a; 2000:
hereinafter K99 and K00) 
L dwarf samples are drawn from magnitude-limited catalogues. As pointed out 
by \"Opik (1924) and Branch (1976), the larger effective sampling volume leads to
enhanced numbers of equal-luminosity binaries under such circumstances. 

We are currently undertaking a project which aims at a definitive measurement of the
binary frequency amongst ultracool dwarfs (spectral types $\ge$M8) by combining  high-resolution
imaging with the Planetary Camera of the Hubble Space Telescope and high-resolution
spectroscopy with NIRSPEC on Keck. This paper presents the first results from
this programme: HST images of twenty L dwarfs, four of which are resolved as binaries. Unlike most
previous discoveries, one of the systems discussed in this paper has a secondary component
significantly fainter than the system primary. That system therefore has a
mass ratio, ${M_2 \over M_1}$ (or $q$), less than unity. 
The following section presents the observations and
the final section discusses the implications of these results. 

\section {Observations}

Table 1 lists the L dwarfs targeted for observation. All 
were identified based on their extremely red near-infrared colours ( (J-K$_S) > 1.3$ mag.) 
measured in the 2MASS survey, and each has been confirmed as spectral type L based on follow-up 
spectroscopy.  Optical spectra of 2M0850, 2M0913, 2M1146, 2M1155, 2M1328, 2M1439
and 2M1632 
(we adopt this abbreviated nomenclature for each source) are presented by K99; 
observations of 2M0036, 2M0746 and 2M1507 are discussed by Reid {\sl et al.} (2000); 
and the remaining L dwarfs are included in K00. Four of the targets have lithium
absorption: 2M0825, 2M0850, 2M1146 and 2M1726  have features with
equivalent widths of 10, 15, 5 and 6\AA\ respectively, 
indicating masses below 0.06 M$_\odot$
(Rebolo {\sl et al.}, 1992). The available spectroscopic observations allow us to set upper limits
of 1\AA\ on Li 6708\AA\ for fourteen of the sixteen remaining dwarfs; low signal-to-noise data
for 2M0708 and 2M1623 lead to upper limits of only 5\AA. All  
are likely to have masses close to or below the hydrogen-burning limit. 

Each L dwarf was imaged on the Planetary Camera chip of WFPC2, using both 
F814W and F606W filters. The camera has a plate-scale of 0.0455 arcsec pix$^{-1}$
and the F814W (I-band) exposure times were adjusted to provide the maximum
dynamic range for companion detection without saturating the target. 
Table 2 gives the journal of observations. The analysis
techniques used are discussed in Reid \& Gizis (1997a, b). In brief, our observations
are capable of detecting equal-luminosity binaries with separations, $\Delta$, of more than
0.09 arcseconds, with limiting sensitivities of $\Delta I_{B-A} = 1$, 3 and 5 magnitudes
at $\Delta = 0.14, 0.23$ and 0.31 arcseconds, respectively, where $\Delta I_{B-A}$ 
is the magnitude difference between secondary and primary. The maximum radius for
companion detection, 18.2 arcseconds, is set by the angular field of view of the PC chip.
The 2MASS scans allow us to search for L- and T-dwarf companions brighter than J=16 at
wider separations. At the average distance of the present sample, 
20 parsecs, equal-luminosity binaries are detectable at separations between
1.8 and 360 AU.

Accurate positions for each source (from 2MASS) are given in Table 1, where we 
also list  spectral types, parallax estimates and
both I-band and infrared photometry. The trigonometric parallax measurements are
from USNO observations (Dahn et al, 1999), updated in a few cases to include
more recent observations (Dahn,  priv. comm.). 
Only nine L dwarfs, including three binaries, have ground-based I$_C$
photometry from USNO observations (Dahn {\sl et al.}, 1999). However, we have used DAOPHOT to
measure I$_{814}$ magnitudes for all of the sources, adopting the zeropoint derived
in Holtzman {\sl et al.}'s (1995) calibration. Clearly, all of our targets are significantly redder
than any of the calibrating stars observed by Holtzman {\sl et al.}, and we have used the
six apparently single dwarfs in this sample to determine the extent of any colour term.
Figure 1 shows the results: the available data can be represented by a
linear correction
\begin{displaymath}
\delta I = I_C - I_{814} = -(0.534\pm0.439) + (0.230\pm0.121) ( I_{814} - J)
\end{displaymath}
Neither the slope nor the zeropoint of the fit is strongly constrained.
However, the dispersion about the relation is only 0.06 magnitudes, and eliminating
the single most discrepant point (at $\delta I = 0.44$ mag) gives 
\begin{displaymath}
\delta I = I_C - I_{814} = -(0.285\pm0.271) + (0.155\pm0.076) ( I_{814} - J)
\end{displaymath}
reducing $\delta I$ by only 0.05 magnitudes at (I$_{814}$-J) = 4. These uncertainties
are not important in the present context.
The increase in $\delta I$ at later spectral types 
probably reflects the steep spectral slope at $\lambda < 9000$\AA\ due
to broad K I $\lambda$7665/7699 absorption. 
We have used the steeper relation to transform the WFPC2 I$_{814}$ magnitudes to the Cousins I-band. 
(Hereafter, we use M$_I$ to denote absolute magnitudes on the Cousins' system, and M$_{814}$
for the HST system.)
Figure 2 plots the location of our targets on the (M$_J$, (I$_C$-J)) plane.

The linear colour term we derive above is unlikely to be valid at  redder colours.
Gl 229B is the only T dwarf with HST photometry. Golimowski {\sl et al.} (1998) measure
M$_{814} = 20.76$, giving an (I$_{814}$ - J) colour of 5.26 magnitudes (adopting the
J-band magnitude measured by Leggett {\sl et al.}, 1999) and predicting
$\delta I = 0.68$ magnitudes. Matthews {\sl et al.} (1996)
list the absolute magnitude as M$_i$=21.2. While this measurement
was made with a Gunn i filter, the magnitude
is tied to the Cousins flux zeropoint, rather than to the AB magnitude system.
Hence, the corresponding colour is (I$_C$-J)=5.68, equivalent to
$\delta I \sim 0.42$ magnitudes.  

Sixteen of the twenty targets in the present sample show no evidence for duplicity in our observations; 
four dwarfs, however, have apparent companions at small angular separations (Figure 3).
Ground-based near-infrared
imaging had already identified 2M1146 as a likely binary source (Ko99). Our
HST imaging confirms that observation, and, using DAOPHOT, 
we measure a separation of 0.29 arcseconds at a position
angle of 199$^o.5$,  and a
magnitude difference of $\Delta I_{814} = I_{814}(B) - I_{814}(A) = 0.31$ magnitudes.
This compares with $\Delta = 0.29$ arcseconds, $\theta = 206^o$ and
$\Delta K = 0.1$ magnitudes derived from NIRC observations (Ko99).  

Reid {\sl et al.} (2000) suggested 2M0746 as a binary candidate based on
its location in the (M$_J$, (I-J)) colour-magnitude diagram, 0.7 magnitudes 
brighter than the ``main sequence''. This suggestion proves to have merit, with
the two components having $\Delta I_{814} = 0.62$ mag
and a separation of 0.22 arcseconds. 2M0920 is barely resolved, but is clearly
extended in the F814W image. Fitting the image profile using our 2M0036
observation as a PSF template, we derive $\Delta I_{814} = 0.4$ mag 
and $\Delta = 0.07$ arcseconds. 

The fourth binary candidate, 2M0850, is notable in that the resolved
companion, lying at a separation of 0.16 arcseconds, is significantly 
fainter than the primary. We measure a
relative magnitude of $\Delta I_{814} = 1.27$ magnitudes. The 
individual magnitudes, based on Holtzman {\sl et al.}'s (1995) calibration, are I$_{814}$ = 20.44
and 21.76 magnitudes, and the fainter source is not detected in the F606W
observation. 

Are these apparent companions physically associated with the L dwarf targets? 
As Table 2 shows, all four candidate binaries lie at moderate to high Galactic latitude, 
with a correspondingly low surface density of background stars. 
There is no evidence that any of the companions is spatially extended, as 
might be expected for background galaxies, although limits are less stringent for 2M0920. 
We have used standard  filtering techniques to remove cosmic rays from
the images, and Table 2 lists the number of faint (I$_{814} < 20.5$), stellar 
sources in each 
Planetary Camera frame. The average surface density is $6.5\pm3.6$ sources per PC frame, implying 
an {\sl a priori} probability of $\sim 3 \times 10^{-4}$ of finding a source within 0.16
arcseconds of a given point on the frame. Even in the more crowded field of 2M0920, the
probability of association, given a random distribution of background sources, is only
$4 \times 10^{-3}$.

Coincidence arguments are vulnerable to small number statistics.
However, faint background sources
are expected to be either K-type Galactic stars or low-redshift ($z < 0.5$)
galaxies with similar colours, and all of the background sources we detect 
have neutral (R$_{606}$ - I$_{814}$) colours.  In contrast, all
of the potential companions have (R-I) colours similar to the known L dwarf
primaries, significantly redder than either K-type stars or galaxies. As a final test, 
Koerner \& Kirkpatrick have deep near-infrared images (K$<22$ mag.) of all four sources, obtained with
the Keck telescope 1 to 2 years before the HST observations. 2M1146A/B is confirmed as a
common proper-motion pair.  The motions of 2M0746 and 
2M0850 (0.38 and 0.16 arcsec yr$^{-1}$ respectively, Dahn {\sl et al.}, 1999) are sufficient 
that the companions should be resolved if either were a stationary background source: no 
such object is visible in either set
of NIRC images. Thus, the preponderance of the evidence indicates that all four faint
sources are physical companions of the respective L dwarf targets.

\section {Discussion}

\subsection{Absolute magnitudes and luminosities}

Adopting the hypothesis that the sources are physically associated, Table 3 lists
the absolute magnitudes derived for the individual components. We have used the
relative fluxes measured from the WFPC2 F814W images to deconvolve the 
relative contribution of each star to the joint I-band photometry. We use
a similar technique to estimate the individual J magnitudes.  The L dwarf
sequence has a slope of $\sim4$ in the (M$_I$, (I$_C$-J)) colour-magnitude plane (Figure 4), 
and we estimate the relative magnitudes at 1.25$\mu$m using
\begin{displaymath}
\Delta J \approx 0.75 \times \Delta I_C 
\end{displaymath}
Absolute magnitudes for each component can then be derived from M$_J$(AB).

Figure 4 plots the location of each component in the (M$_I$, (I$_C$-J)) plane. The
two most luminous systems, 2M0746 and 2M1146, are both  overluminous in
Figure 2, which plots the joint, ground-based photometry. 
Figure 4 shows that the individual components of both
systems lie squarely within the L dwarf sequence\footnote{ Two unresolved systems
are noticeably over luminous in Figure 4: PC0025, 
at M$_I = 15.5$, (I$_C$-J)=3.7; and 2M1328, at  M$_I = 17.2$, (I$_C$-J)=4.1.
Burgasser {\sl et al.} (2000a) suggest that the former is an interacting red dwarf/brown dwarf
binary. Both would repay further study.}.
These are two of the brightest L dwarfs known; indeed,
2M0746 is currently the brightest, with an apparent magnitude of K=10.49. The fact
that both prove to be binaries, as have two of the three bright L dwarfs
discovered in the initial DENIS brown dwarf minisurvey (Delfosse {\sl et al.}, 1997), 
is a clear example of \"Opik's  equal-mass binary selection effect. 

Considering the two later-type binaries, both components in 2M0920 lie
0.3 magnitudes blueward of the L dwarf sequence. That system currently
lacks a trigonometric parallax, and the absolute magnitudes are correspondingly 
less certain. 
Both components in 2M0850AB lie squarely on the
sequence. There is something of a discrepancy between the observed spectral
type, L6, and the absolute magnitude of 2M0850A, which is slightly fainter
than the value derived for the L7.5 dwarf, 2M0825. This may simply
reflect intrinsic scatter in the (M$_I$, spectral type) relation. 

Given these observations, we have attempted the derivation of luminosities
of these systems. In principle, theoretical models can be used for this purpose.
However, the flux distribution of ultracool dwarfs is affected
by atmospheric dust, which starts to form at approximately spectral type M6. The
emergent energy distribution predicted by stellar models is dependent strongly on
the assumption made concerning the distribution of that dust, as illustrated most
dramatically by Chabrier {\sl et al.} (2000): dust-free and dusty models differ by over 2
magnitudes in (J-K$_S$); bolometric corrections have correspondingly large  uncertainties.
These problems are particularly acute for L-type dwarfs.
Under such circumstances, we prefer to rely on empirical
measurements.

Even so, it is clear that there is room for improvement in the definition of
empirical bolometric corrections for  late-type dwarfs. 
Figure 5 illustrates the current state of knowledge, plotting the available J-band and I$_C$
bolometric corrections. While we have reliable measurements for Gl 229B (from Leggett {\sl et al.}, 
1999), the latest type L dwarf with a bolometric magnitude estimate is the L4 dwarf GD 165B
(Jones {\sl et al.}, 1994; Kirkpatrick {\sl et al.}, 1999b). Figure 5 plots those data,
together with absolute magnitudes and bolometric corrections for the L2 dwarf, Kelu 1
(Ruiz {\sl et al.}, 1997), and the M9.5 dwarf, BRI0021-0021 (Tinney {\sl et al.}, 1993). 
The arrows in the upper
panel for Figure 5 mark the observed locations of 2M0850A and 2M0850B.

We estimate M$_{bol}$ for the binary components using two techniques: first, we
use linear interpolation in the (M$_I$, BC$_I$) diagram. This approach may
overestimate BC$_I$ (M$_{bol}$ too bright) in later-type L dwarfs, since the
bolometric correction must depend strongly on the extent of KI absorption, which
is unlikely to increase in such a simple manner from L4 to T. Second, we combine
the M$_J$ values listed in Table 3 with an estimated BC$_J$=1.9 magnitudes (lower
panel, Figure 5). 
Averaging those estimates, which generally agree to within 0.3
magnitudes, gives the values of $\langle$M$_{bol} \rangle$ listed in Table 3.
Luminosities are calculated for an adopted value of M$_{bol\odot} = 4.72$.

2M0850B stands out as particularly intriguing object.  The I-band  absolute
magnitude is  $\sim0.9$ magnitudes  fainter than the L8 dwarf 2M1632 (Table 1), while
the estimated J-band absolute magnitude, M$_J \sim 15.2$, is 0.2
magnitudes fainter than faintest L dwarf currently known, Gl 584C (K00), and
only 0.3 magnitudes brighter than the value measured for Gl 229B (Leggett {\sl et al.}, 1999). 
Given these absolute magnitudes, 
the companion could be either a late-type L dwarf, spectral type $\approx$L9,
or an early-type T dwarf.
We discuss this system in more detail in section 3.3. 

\subsection {Masses and temperatures}

Our HST observations provide direct measurement of the relative magnitudes of the
components; transforming
those data to estimates of effective temperature and mass is not straightforward. 
Brown dwarfs evolve rapidly in luminosity and temperature as a function of mass. 
The components of a binary system, however,  can be assumed to be coeval. Hence, the difference
in luminosity observed for 2M0850A/B must be interpreted as a 
difference in mass. This is in contrast to most other known L dwarf binaries, 
where both components have near-equal luminosity
and hence near-equal mass. L dwarfs and T dwarfs are known as secondary components of
main-sequence stars (eg Gl 229B, Nakajima {\sl et al.}, 1995; G196-3, Rebolo {\sl et al.}, 1998), but
at much wider separations (30 to 4000 AU). 2M0850A/B provides 
direct evidence that unequal-mass brown dwarf binaries can form at separations of
$<10$ AU. 

Evolutionary tracks for brown dwarfs of different masses lie in close proximity in the HR diagram, 
overlapping tracks for low-mass stars at higher luminosities. As a result, 
the mass-luminosity relation is not single-valued in this r\'egime,  and we cannot estimate reliable
masses for individual objects without knowledge of their age. However, if we
can determine the bolometric magnitude of each component in a brown dwarf binary, we can use
models to estimate a mass ratio based on the relative luminosity. 

Given the luminosity estimates in Table 3,  we have used theoretical models
computed by Burrows {\sl et al.} (1993, 1997) to constrain the values of relative masses
in the lower luminosity systems, 2M0920 and 2M0850. As noted above, the
presence of lithium in the optical spectrum of 2M0850 sets an upper limit of
0.06 M$_\odot$ on the mass of the primary component in that system.  
Both components of 2M0920AB have a luminosity similar to the value
for 2M0850A, but spectroscopic observations set an upper limit of
0.5\AA\ for the equivalent width of the Li 6708\AA\ line. Theoretical
models predict that lithium is removed from the gas phase as solid LiCl at low
temperatures (Fegley \& Lodders, 1996; Burrows \& Sharp, 1999; Lodders, 1999), 
and spectroscopy
of L dwarfs indicates that the strength of Li 6708\AA\, when detected, decreases
amongst the latest spectral types (K00). However, $\sim 50\%$ of L6/L6.5 dwarfs have
detectable lithium, so it is more reasonable to assume that the absence of
lithium in 2M0920 is due to destruction through fusion, rather than solidification. In
that case, both components in 2M0920AB have masses exceeding 0.06$M_\odot$. 

Accepting these hypotheses, Figure 6a superimposes the luminosity
estimated for each of the four 
binary components on the (log(L), log(age)) plane. 
Figure 6b plots the corresponding mass ratios, $q$. In both cases,
the latter parameter varies over a relatively small range, with $q \sim 0.96$ for
2M0920AB and $q \sim 0.75$ for 2M0850AB. This sets  an upper limit of 0.05 M$_\odot$
 for the mass of 2M0850B.

The Burrows {\sl et al.} (1993, 1997) models also
allow us  to estimate approximate temperatures for each component,
given a particular value for the primary mass (i.e. age).
Table 4 lists those values, where we have interpolated between tracks as necessary.
We note that there are some inconsistencies, particularly at low masses/young ages: for
example, if 2M0850A has M$\sim 0.02 M_\odot$, Figure 6a implies 
$\tau \sim 0.19$ Gyrs and $\sim 0.015 M_\odot$  for 2M0850B; a 0.015M$_\odot$ brown
dwarf is predicted to have a log(${L \over L_\odot}) \sim -4.7$ at that age, however, rather than 
the observed log(${L \over L_\odot}) \sim -4.9$. Nonetheless, these calculations provide
a preliminary indication of the properties of the components in each system.

Figure 7 illustrates the results, superimposing 
 on the theoretical H-R diagram the error boxes for both components of 2M0850 and
for 2M0920A. The similarity in spectral types between the composite spectra of the
two systems suggests that 
2M0850A lies near the upper limit of the estimated mass range. The most probable masses for the
individual components are therefore: \\
\centerline {2M0850A: 0.05$\pm0.01$ M$_\odot$} \\
\centerline {2M0850B: 0.04$\pm0.01$ M$_\odot$} \\
\centerline {2M0920A: 0.068$\pm0.008$ M$_\odot$} \\
\centerline {2M0920B: 0.068$\pm0.008$ M$_\odot$} 

There are fewer constraints on the masses of the components in the two earlier-type
binaries. As with 2M0850, lithium absorption in 2M1146 indicates masses below 0.06$M_\odot$
for both primary and secondary. 
2M0746, in contrast, is lithium free, and if the system is older
than $\sim1$ Gyr, the Burrows {\sl et al.} models predict that both components exceed the 
hydrogen-burning mass limit; at age 10 Gyrs, both have masses of 0.085$M_\odot$.

Reliable masses require orbit determination. Such measurements should
be possible over a reasonable timescale for at least two of the four systems
discussed here. The average observed separation of components in a binary system 
is eighty percent of the semi-major axis. Applying that rule of thumb to our
observations, and taking the mass limits estimated above, 2M0850AB is likely
to have an orbital period of 43 years, while 2M0920AB may have a period as short as 8.5 years. 
The latter system may be amenable to astrometry with the next generation of optical
interferometers. 2M0746 may be accessible to more conventional imaging: with
a lower mass limit of 0.12 $M_\odot$ (total mass), a semi-major axis of 3.5 AU implies
a period of less than 18 years. In contrast, 2M1146 ($M_{tot} < 0.12 M_\odot$) with
$\langle a \rangle \sim 9.5$ AU has a likely period exceeding 85 years.

\subsection {Temperature scales, 2M0850 and the L/T transition}

The extensive observational and theoretical work undertaken over the last few
years has led to the emergence of a convincing qualitative scenario for
brown dwarf spectral evolution. As the atmosphere cools below an effective
temperature of $\sim2500$K, first TiO and then VO solidify as dust particles,
leaving metal hydrides as the strongest molecular features in the optical
spectrum. The removal of these major opacity sources leads to increased
atmospheric transparency, which in turn accounts for the presence of strong,
pressure-broadened atomic lines of alkali metals (Na, K, Cs, Rb, Li). At
near-infrared wavelengths, H$_2$O absorption dominates the spectrum, with
CO a prominent feature at 2.03$\mu$m. As originally predicted by Tsuji (1964), 
carbon is bound preferentially in methane, rather than CO,  at low temperatures, leading to 
strong absorption in the H and K windows, as in Gl 229B. This accounts for the
blue near-infrared colours of T dwarfs such as Gl 229B and Gl 570D.

The qualitative picture is clear. Quantitatively, however, we still lack a well-grounded
temperature scale to associate with this behaviour. 
The threshold temperature (if, indeed, there is a clean threshold)
for the onset of methane absorption at 2$\mu$m is
important not only in understanding the structure of
individual brown dwarfs, but also in disentangling the underlying mass function
from the observed number/spectral type surface densities (Reid {\sl et al.}, 1999). 
In broad terms, two temperature calibrations have been suggested. 
Basri {\sl et al.} (2000) have
combined model atmosphere calculations with high-resolution line profiles to
derive temperatures of $\sim 2200$K for spectral type L0 and $\sim 1700$K
for type L8 (K99 classification system). Noll {\sl et al.} (2000) argue that the
relatively weak 3.3 $\mu$m fundamental-band methane absorption detected in mid- and
late-type L dwarfs favours this relatively hot scale, which implies a $\sim700$K
temperature difference between the latest L dwarfs and Gl 229B. If this
temperature scale is correct, then there is a substantial population of
early-type T dwarfs in the Solar Neighbourhood, at least equal in number density to
the known L dwarf population.

Photometric analyses, however, suggest lower temperatures. 
At least four currently-known L8 dwarfs have  
values of M$_J$ within 1 magnitude of Gl 229B (M$_J$ = 15.5) : 2M1632, M$_J$ = 14.7
(Table 1), SDSSp J132629.82-003831.5 (M$_J$=14.8; Fan {\sl et al.}, 2000), 2MASSW J0310599+164816
(M$_J = 14.9$, K00) and 2MASSW J1523226+301456 (M$_J = 15.0$, K00). The last mentioned dwarf is
particularly important, since it is a proper-motion companion of the known nearby
G-dwarf binary, Gl 584AB, and therefore has a well-determined trigonometric parallax.  

Late L dwarfs and T dwarfs have radically different colours: L8 dwarfs have (I-J)$\approx$4.2,
(J-K)$\approx$2, as compared with (I-J)$\approx$5.7 and (J-K)$\approx$0 for Gl 229B.
However, those differences are tied strongly to the relative strength of individual absorption
features, broad KI at I and CH$_4$ overtone bands at H and K, rather than 
substantial variation in the underlying energy distribution.
In contrast, the J-band is largely free of significant absorption features: the
crucial question is how well the flux in that passband tracks bolometric
magnitude. Is ${F_J \over F_{bol}} \approx$constant, as Figure 5 suggests?
The scarcity of mid-infrared data for ultracool dwarfs means that
we cannot answer this question directly at present; however, we can examine some
of the consequent eventualities. 

Gl 229B, the original T dwarf, has an effective temperature of $960\pm70$K 
(Marley {\sl et al.}, 1996). 
Since brown dwarf radii are set by degeneracy, we know that 2M1523 (Gl 584C) and Gl229B
have radii which differ by at most 15\%. We have
\begin{displaymath}
{L_1 \over L_2} = ({R_1 \over R_2})^2 \ ({T_1 \over T_2})^4
\end{displaymath}
Suppose that Gl 584C is a high-mass brown dwarf, with a radius 15\% smaller than that
of Gl 229B. 
In that case, if Gl 584C has T$_{eff}=1700$K, then L$_{Gl 584C} \sim 5.5 L_{Gl 229B}$,
or M$_{bol} = 15.5$, and the J-band bolometric correction is only -0.5 magnitudes.
On the other hand, if we assume that the bolometric corrections outlined in Figure 5 are
reliable, then L$_{Gl 584C} \sim 2 L_{Gl 229B}$. In that case, assuming a 15\%
difference in radius, Gl 584C is predicted to have a temperature 30\% higher than that observed
for Gl 229B. That is, the photometric temperature estimate for spectral
type L8 is $1250\pm100$K (K00), $\sim250$K cooler than Basri {\sl et al.}'s  spectroscopic estimate.

Figure 2 underlines visually the photometric hypothesis, which is essentially a
continuity argument.  The L dwarf sequence spans a
three magnitude range in the J band, $12 \le M_J \le 15$; Basri {\sl et al.} assign
a corresponding temperature range of 500K. Under this scenario, the 0.5 magnitude 
range $15 \le M_J \le 15.5$ between Gl 584C and Gl 229B corresponds to a temperature
difference of 700K. The corresponding photometric calibration assigns a
temperature difference of $\sim750$K (2000 to 1250K) to the L dwarf sequence, and 
$\sim250$K to the latter interval.

The discovery by the SDSS collaboration of three field brown dwarfs with both CO and CH$_4$ 
absorption (Leggett {\sl et al.}, 2000) will clearly contribute
valuable information on this problem.
The co-existence of CO and methane in so many systems is surprising, since
theoretical models predict their co-existence over a relatively narrow range in
temperature  (Fegley \& Lodders, 1996). This apparent paradox might be resolved
through a combination of two factors: first, binarity (as discussed further below); 
and, second,  the non-grey nature of late-L/T-dwarf atmospheres. As with
Jupiter, topographical structures (bands, spots, zones) may allow us  
to see to different depths at different physical locations: that is, the overall energy
distribution is produced by a range of temperatures, rather than being characterised by
a single effective temperature, as is usually taken to be the case in hotter stellar atmospheres. 
For example, the 3.3 $\mu$m methane fundamental  band 
is believed to be produced at higher levels in the atmosphere (lower pressure,
lower temperature) than are responsible for the ``continuum'' flux in late L dwarfs. 
Under such inhomogeneous conditions, both CO and the overtone 1.6 and 2.1 $\mu$m CH$_4$ bands 
might be detectable in brown dwarfs spanning  a broader range in luminosity than expected. 

As yet, none of the SDSS T dwarfs has a known luminosity.
2M0850AB has a measured trigonometric parallax, allowing us to locate both
components on the HR diagram (Figure 4). Thus, this system, and others like
it, offer the prospect of providing valuable insight into this matter by
linking the unknown (2M0850B) with the barely known (2M0850A).

The onset of detectable methane absorption in the K window defines the
transition from spectral class L to class T (K99). Photometrically, 
however, we must make a distinction between cool dwarfs like Gl 229B, with near-saturated
methane absorption in the H and K bands and A-type near-infrared colours, and
the objects discovered recently by Leggett {\sl et al.} The latter are, by definition, 
T dwarfs, since CH$_4$ is evident at K, but their near-infrared colours are closer to
those of M giants, reflecting the weaker CH$_4$ absorption. We will refer to these 
objects as early-type T dwarfs, and describe the more familiar Gl 229B-like
objects as `classical' T dwarfs.

Given its absolute magnitude, 2M0850B might be a very late-type L dwarf, an early-type T dwarf
or a `classical' T dwarf.
High spatial-resolution near-infrared photometry would allow us
to discriminate amongst these options, since each occupies a distinct location in
the (J-H)/(H-K) plane. Currently, we lack such data, but the ground-based photometry
listed in Table 1, together with K-band spectroscopy (K99),
allow us to set limits on the nature of the companion. Based on the slope of
the (M$_I$, (I-J)) L dwarf sequence, we inferred $\Delta J_{B-A} \sim 1$ magnitude, which
implies that 2M0850B supplies $\sim30\%$ of the flux at 1.25$\mu$m in joint photometry.
Late-type L dwarfs have similar (J-K) colours, so one would expect a similar contribution at K
in an L-dwarf/L-dwarf binary system. In contrast, Gl 229B loses $\sim70\%$ of its K-band
flux to methane absorption, so, given the same  $\Delta J_{B-A}$, 
an L-dwarf/classical T-dwarf system would be $\sim0.2$ magnitudes
bluer in (J-K) than an L-dwarf/L-dwarf binary. L-dwarf/early T-dwarf binaries will
have intermediate (J-K) colours. 

2M0850AB has a (J-K) colour of $1.85\pm0.06$ magnitudes. In comparison, 2M0825
and 2M1632, the two apparently-single L dwarfs closest to 2M0850A in M$_I$,
have (J-K) colours of $1.97\pm0.10$ and $1.86\pm0.05$ magnitudes, respectively. 
All three dwarfs have photometry from the same source (USNO), so this comparison
indicates that 2M0850 is not unusually blue for its spectral type and
absolute magnitude. 

We have modelled the spectrum of a hypothetical late-L/T-dwarf binary
by combining our CGS4 observations of the L7 dwarf, DENIS-P 0205.4-1159
(Reid {\sl et al.}, in prep.), with similar data for Gl 229B (Geballe {\sl et al.}, 1996).
Using IRAF, the data were binned to same scale (5\AA\ per pixel) and summed,
adopting $\Delta J_{B-A} = 1$ magnitude. The combined spectrum is plotted in
Figure 8, and compared with our K-band data for 2M0850AB (K99). 
The J-band magnitude difference we infer for the components corresponds
to $\Delta H_{B-A} \sim 2$ magnitudes and $\Delta K_{B-A} \sim 2.6$ magnitudes. 
However, the extensive methane absorption in a Gl 229B-like T dwarf
leads to a significantly smaller flux difference
in the shorter wavelength half of both H and K  passbands, and the companion 
distorts the composite spectral energy distribution at those wavelengths, 
particularly in the H band. 

Figure 8 shows that the
2M0850AB K-band observations are a reasonable match to our hypothetical composite spectrum
at $\lambda > 2.15$ $\mu$m, but fall below the predicted flux distribution 
at shorter wavelengths. A T dwarf companion with strong CH$_4$ absorption would be 
expected to make the most significant contribution in the latter spectral region.
Thus, the observed discrepancy argues against that option.

There is a noticeable similarity between the near-infrared spectrum of 
our hypothetical late-L/T-dwarf binary and data for the early-type T dwarfs
identified by Leggett {\sl et al.} (2000). Indeed, following \"Opik's binary-selection
criterion, one might expect the brighter `early T-dwarfs' to include a number
of unresolved near equal-mass L/T binaries, with relatively small magnitude
differences $\Delta J_{B-A}$ and $\Delta z_{B-A}$. 
Distinguishing those objects 
from isolated early T dwarfs should be possible by examining the strength of 
the weaker methane absorption bands (for example, the 2$\nu_2$ 1.67 $\mu$m
feature) in both H and K bands. Broadband colours may also provide discrimination:
combining a late-type L dwarf, (J-K)$\sim 1.9$, with a T dwarf companion, 
$\Delta J_{B-A} = 0.0$, $\Delta K_{B-A} = 1.9$, will produce a joint colour
of (J-K)$\sim 1.3$ magnitudes. Bluer colours, such as those observed for three of
the SDSS early-type T dwarfs, can only be produced if the companion T dwarf is
{\sl brighter} than the late-type L dwarf in the J passband. Given the relatively
low opacities in the J passband and the extremely non-grey flux distribution at
these low temperatures,  such circumstances cannot be excluded.

Summarising the discussion, 
both broadband photometry and the observed K-band spectrum suggest that 2M0850B 
is unlikely to be a `classical' T dwarf,  with full blown CH$_4$ absorption. It remains possible
that the companion is an early-type T-dwarf. Near-infrared photometry or
spectroscopy of the individual components (or perhaps joint spectroscopy at H)
should provide a definitive answer to this issue. 

\section {L dwarfs in binary systems}

The ultimate goal of our program is a determination of the frequency 
of binary and multiple  L dwarf systems.  With observations of only twenty targets, our 
current conclusions must be tentative. Moreover, the present observing list
is far from the complete,
volume-limited sample which is ideally suited to this type of investigation. 
However, our observations are not limited to the brightest known L dwarfs, and
are therefore less subject to \"Opik's equal-mass binary selection effect. Bearing
these caveats in mind,  our    initial results suggest that
the frequency of wide binaries amongst L dwarfs is {\sl lower} than that observed
for M dwarf stars. 

\subsection {The HST sample}

Figure 9 compares  the observed magnitude difference, $\Delta I_{814}$, for the four
binaries detected in this sample against the formal detection limits for
HST Planetary Camera observations (Reid \& Gizis, 1997b). Those limits may
be overly pessimistic: 2M0920 lies
slightly below those formal limits, while 2M0850AB is easily detected, despite
its proximity to the detection limit at that separation. At larger separations, the
2MASS data allow us to exclude any L dwarf companions with J$< 16$ (typically M$_J < 14.5$, 
M$_I < 18$) and $\Delta < 10$ arcminutes.  It is clear that
the detected systems occupy only a small fraction of the total parameter space 
available for the detection of potential companions. 

The frequency of resolved binary L dwarfs in the present sample is $20\pm11\%$.
We can match this statistic against results from three other high-resolution
imaging surveys: Reid \& Gizis (1997b) obtained WFPC2 images of 53 late-type M dwarfs in the Hyades cluster,
identifying nine ($17\pm7\%$) confirmed binaries; similarly, 
WFPC2 observations of 41 field M dwarfs resolve 8 ($19.5\pm7.5\%$) as binary or multiple systems (Reid \&
Gizis, 1997a); finally, none of the 27 Pleiades very low-mass (VLM) dwarfs 
surveyed by  Mart\'in {\sl et al.} (2000) with  the NICMOS 1 camera have resolved companions. 
The last sample includes targets with masses comparable with those expected for our field L dwarfs,
while the Hyades and field  M dwarfs have higher masses, with
approximate limits of $0.1 < {M \over M_\odot} < 0.3$
and $0.2 < {M \over M_\odot} < 0.5$, respectively.

Each of these four samples has a different mean distance. We take this into
account by transforming the results to the linear r\'egime. Taking the sensitivity limits
plotted in Figure 9 as a template, we calculate the limiting absolute magnitude for
companion detection as a function of separation, $\Delta$, in AU; combining the 
results for each target allows us to estimate completeness limits for each sample, 
that is, the fraction of targets where we would expect to detect a companion with
a given absolute magnitude at a given linear separation. 
Figure 10 plots those results, where we show the 50\% and 100\% detection limits
together with the detected companions and the approximate absolute magnitude range 
of the targets.

Simple visual comparison of the four panels in Figure 10 suggests a
significant difference in the semi-major axis distribution of the M dwarf and L dwarf
binaries: the L dwarf binaries all have separations of less than 10 AU, while the
overwhelming majority of the M dwarf companions have $\Delta > 10$ AU.
Indeed, were the L dwarfs to lie at the distance of the Hyades, we could expect to
resolve only one system, 2M1146, while none would be resolved at the distance of the Pleiades. 
Mart\'in {\sl et al.} identify six Pleiades dwarfs (including PPl 15) as candidate photometric
binaries. That fraction, $22\pm9$\%, is consistent with the statistics of our L dwarf sample, 
where we resolve four systems and identify 2M1328 as a candidate photometric binary.

Conversely, the wider binaries present in the 
M dwarf samples are not present either in our L dwarf sample or, as already
noted by Mart\'in {\sl et al.} (2000), amongst the Pleiades VLM dwarfs. Fourteen percent
of the Hyades and field M dwarfs have companions at separations between 10
and 100 AU, with a further 5\% having companions in the range $100 < \Delta < 1000$ AU.
A similar distribution pertains for stars in the immediate Solar Neighbourhood, with 
binary fractions of 5\% at $\Delta  \le 1$ AU; 18\% at $1 < \Delta \le 10$ AU
(compatible with our L dwarf sample); 
11\% at  $10 < \Delta \le 100$ AU; and 6\% at  $100 < \Delta \le 1000$ AU
(Reid \& Gizis, 1997a). Combining our HST observations
with the 2MASS limits, we would expect 3 to 4 wide ($\Delta > 10$ AU) binaries in the current
sample:  Mart\'in {\sl et al.} identified similar expectations in their analysis
of the Pleiades sample. The absence of such binaries from both samples implies a
3$\sigma$ deficit of wide L-dwarf binaries.

\subsection { The semi-major axis distribution of brown dwarf binaries}

Our HST observations are aimed at identifying low-mass binary systems. However, both
L dwarfs and T dwarfs are found as companions to higher-mass main sequence stars.
We have therefore made a preliminary effort to set the current results in a broader
context.

Table 5 collects the available data for binary and
multiple systems which include one or more ultracool dwarf component(s). With the
exception of HD 10697B and the Pleiades brown dwarf, PPl 15AB, mass estimates are based
on either the detection/non-detection of lithium or the estimated age of the system
(usually based on the level of chromospheric activity exhibited by the
main-sequence primary star).
At present, the selection effects which underlie this sample are
too diverse to allow detailed statistical analysis. However, two qualitative
comments can be made: 
\begin{itemize}
\item Radial velocity surveys, which have identified more than forty planetary-mass
companions of nearby G dwarfs, have discovered a bare handful of
brown dwarf companions with orbital semi-major axes $a < 10$ AU. 
This is the  `brown dwarf desert', highlighted by Marcy \& Butler 
(1998).  Table 5 shows that brown dwarfs exist as {\sl wide} (10 to 4000 AU)
companions of nearby solar-type stars. Systems with comparable separations, but
with an M dwarf as the wide component, are also known, notably
Proxima Cen/$\alpha$ Cen (Matthews \& Gilmore, 1993). 
\item In contrast, all L-dwarf/L-dwarf binaries detected to date have
separations of less than 10 AU, overlapping with the `brown dwarf desert'
in higher-mass systems, and most have mass ratios near unity.  
Close brown dwarfs clearly form with greater ease around low-mass primaries
(cf. Basri \& Mart\'in, 1999)
\end{itemize}
Figure 11 presents these results in graphical form. The upper panel
compares the ($q, \Delta$) distribution of the the L dwarf binaries
listed in Table 5 against similar data for M dwarf binaries within 8 parsecs of
the Sun (from Reid \& Gizis, 1997a) and for the HST M dwarf samples discussed
in the previous section. Several features of this diagram demand comment:

First, there is an apparent  preference for equal-mass L-dwarf/L-dwarf binaries.
It remains possible that the observed distribution is partly a result of selection
effects, since current observations have only limited sensitivity to
low-$q$ systems at small $\Delta$ (see Figure 10). 
Nonetheless, this echos the bias toward  equal-mass systems amongst
M dwarf binaries (Reid \& Gizis, 1997a, 1997b), where observations
extend to lower mass ratios. 

Second, no L-dwarf/L-dwarf binaries are known with separations $\Delta > 10$AU -
is this nature or nurture? There are several possibilities: wide low-mass
binaries form, but are stripped by gravitational interactions; wide binaries 
are inhibited from forming, leading to a truncated semi-major axis distribution 
and overall decrease in binary frequency; or formation of wide binaries is inhibited, but
close systems form with greater frequency amongst VLM dwarfs. The last option was 
suggested by Basri \& Mart\'in (1999), prompted by the existence of the 
short-period Pleiades binary, PPl 15. Better statistics on the frequency of
such binaries amongst VLM dwarfs can constrain the latter two options. 
We note that Mayor (2000) finds 
that the fraction of spectroscopic (small separation) binaries is
essentially constant at $\sim12\%$ for nearby stars with spectral types between G and M. 

The third outstanding feature of the ($q$, log($\delta$)) diagram is the
prevalence of low-$q$ systems at wide separations. These are 
low-mass analogues of the wide, common proper motion (cpm) systems made familiar by
Luyten's extensive surveys. Such systems are vulnerable to disruption  
through gravitational encounters with
stars and giant molecular clouds (GMCs), or via tidal effects due to the Galactic field. 
The binding energy of a binary star system is directly
proportional to the total mass and inversely proportional to the
separation.  If we make the reasonable assumption that all binaries 
are subject to gravitational perturbations of the same average force, then
we would expect to find fewer wide binaries with decreasing total system mass.
Consequently, low-mass stars and brown dwarfs in wide binaries are more likely to
survive if the primary has a relatively high mass, leading naturally to a
predominance of low-$q$ systems at large $\Delta$.
Indeed, we might expect a characteristic cutoff radius for binary separation 
as a function of total mass, M$_{tot}$. 

The lower panel in Figure 11 suggests strongly that there is a correlation between
maximum separation and M$_{tot}$. Such behaviour would provide a natural explanatiuon for
the decreased {\sl total} binary frequency amongst M dwarfs (as compared with G dwarfs),
while preserving the invariance with mass of the frequency of spectroscopic binaries.
However, it is not clear whether
this follows the functional form expected for tidal disruption. The dotted line
plotted in Figure 11 is 
\begin{displaymath}
log(\Delta_{max}) \ = \ 3.33 M_{tot} \ + \ 1.1
\end{displaymath}
i.e. a log-normal relation (the three M-dwarf binaries beyond the line are
Gl 412AB, RHy 240AB and MT 3AB). One expects a scaling linearly proportional to
$M_{tot}$ for disruption by
point source encounters, although circumstances are more complex in the
case of an extended potential (Weinberg {\sl et al.}, 1987). It seems unlikely, however,
that the observed rapid decrease in $\Delta_{max}$ with M$_{tot}$ is due solely to
dynamical evolution.

As a final comment, we note that 
the wide cpm systems plotted in Figure 11 have dimensions which exceed those
of typical protostellar disks. Previous studies have suggested that the 
relative properties of short and long-period stellar binaries may
differ significantly (Mazeh {\sl et al.}, 1992), perhaps reflecting
different formation histories (Mathieu, 1994).  
Individual components in these wide systems may have formed essentially independently.
Indeed, the widest systems may be more akin to residual cluster fragments 
- that is, progeny of separate cloud cores which retain the motion of the parent cluster. 
Under this scenario, Gl 584ABC, $\alpha$ Cen/Proxima, Gl 752A/VB10 and their ilk would be
minimal examples of Eggen-style moving groups - cousins, rather than siblings. 

Further examples of wide L-dwarf common proper motion systems will be 
forthcoming from detailed examination of the photometric 
catalogues produced by 2MASS, SDSS  and DENIS.
The completion of our present survey will provide
more detailed information on the prevalence of ultracool dwarf binaries at modest separations. 

\section {Summary}

We have presented high spatial-resolution Planetary Camera observations of twenty
L dwarfs. Four are resolved as binary systems, with fainter companions at
projected separations of 2 to 8 AU. While our present sample consists of only
20 L dwarfs, drawn from a magnitude-limited, rather than volume-limited, parent
sample, the preliminary indications are that the fraction of binary L dwarfs at
at separations exceeding 10 AU 
is {\sl lower} than the empirical value of $\sim20\%$ measured for M dwarf systems. 
Both L and T dwarfs, however, are found as wide cpm companions of higher-mass,
main-sequence stars.

Three of the four binaries in our current L dwarf sample
 here have components with similar luminosities,
implying nearly equal masses; 2M0850AB, however, has a secondary component which 
is 1.3 magnitudes fainter than the primary. The primary has 
strong lithium absorption, indicating a mass
below 0.06 M$_\odot$, and comparison with theoretical models calculated by 
Burrows {\sl et al.} (1993, 1997) suggests a secondary/primary mass ratio of $\sim 0.8$. 2M0850B
has M$_I = 19.8\pm0.25$, and is likely to be either a very late-type L dwarf ($\sim$L9) or
an early-type T dwarf.

Unequal-mass brown dwarf binaries  offer an effective means
of constraining theoretical models. Since the components are coeval, both must
lie on the same isochrone in the (log(L), log(T$_{eff}$) plane. This
technique has been used to test models of pre-main sequence stars and brown dwarfs
against observations of young, multiple systems (see, for example, White {\sl et al.}'s (1999)
analysis of data for GG Tau).  
Once accurate effective temperatures are available for L dwarf systems, similar methods
can be used to probe parameter space at lower masses and temperatures, as illustrated in Figure 7.
In particular, this approach offers excellent promise for 
resolving questions concerning the L/T transition phase: determining both the
critical threshold temperature for the onset of the 
the change from CO-dominated spectra (L dwarfs) to CH$_4$ spectra (T dwarfs), and
the rapidity of the transformation. More detailed observations of these, and other,
binary L-dwarf systems will provide benchmarks for future theoretical models of 
low-mass stars and brown dwarfs.

\acknowledgments { } The authors would like to thank the referee, Gibor Basri, for
useful criticism.  This work is  based on observations with the NASA/ESA Hubble Space Telescope, 
obtained at the Space Telescope Science Institute, which is
operated by the Association of Universities for Research in Astronomy, Inc. 
under NASA contract No. NAS5-26555. 
Our research was supported by NASA through ST ScI Grant GO-08146.01-97A.

\newpage

\begin{table}
\begin{center}
{\bf Table 1: Astrometric and photometric data}
\begin{tabular}{lccccccc}
\tableline\tableline
2MASS & Sp. & Ref. &$\pi$ (mas)& M$_I$ & (I-J) & M$_J$ &(J-K) \\
\tableline
WJ0036159+182110 & L3.5 & R00 &112.4$\pm2.0^1$ & 16.36$\pm0.05^3$ & 3.67 &12.69$^3$ & 1.38$\pm 0.03$ \\
WJ0708213+295035 & L5   & K00 & 23$\pm3.5^2$ &17.51$\pm0.35^4$ &3.98 & 13.53 & 2.06$\pm 0.15$ \\
WJ0740096+321203 & L4.5 & K00 & 27$\pm 4^2$ &17.07$\pm0.35^4$ &3.74&13.33 & 1.99$\pm0.11$ \\
WJ0746425+200032 & L0.5 & R00 & 83$\pm 2^1$ & 14.69$\pm0.06^3$ & 3.38 & 11.31$^3$ & 1.24$\pm0.04$ \\
WJ0820299+450031 & L5 & K00 & $28\pm4^2$ &$17.65\pm0.35^4$  &4.13   & 13.52 & $2.06\pm0.14$ \\
WJ0825196+211552 & L7.5 & K00 & 80$\pm12^1$ &18.29$\pm0.30^5$ &4.23 & 14.06$^3$ & 1.97$\pm0.10$ \\
sJ0850359+105716 & L6 & K99 &36.1$\pm4.4^1$ & 18.21$\pm0.25^3$ & 4.31 & 13.90$^3$ &1.85$\pm0.06$ \\
WJ0913032+184150 & L3 & K99 & $29\pm4^2$ & 16.83$\pm0.35^4$ & 3.63 & 13.20 & $1.72\pm0.09$ \\
WJ0920122+351742 & L6.5 & K00 &48$\pm7^2$ & 17.68$\pm0.35^4$ & 3.73 & 13.95& 1.66$\pm0.11$ \\
WJ0928397-160312 & L2 & K00 & $26\pm4^2$ & $15.96\pm0.35^4$  &3.55 &  12.41 & $1.70\pm0.07$ \\
WJ1123556+412228 & L2.5 & K00 & $20\pm3.5^2$ &$16.32\pm0.35^4$ &3.75 & 12.57 & $1.70\pm0.07$ \\
WJ1146345+223053 & L3 & K99 & $38.2\pm1.3^1$ & $15.76\pm0.05^3$ & 3.73 & 12.03$^3$ & $1.53\pm0.04$ \\
WJ1155009+230706 & L4 & K99 &30$\pm4.5^2$ & 17.12$\pm0.35^4$ & 3.73 & 13.39 &1.68$\pm0.15$ \\  
WJ1328550+211449 & L5 & K99 & $26.3\pm4.9^1 $ & $17.19\pm0.35^3$ & 4.12 & 13.07$^3$ &  $1.85\pm0.05$ \\
WJ1338261+414034 & L2.5 & K00 & $44\pm7^2$   &15.89$\pm0.35^4$  &3.45  & 12.44 & $1.47\pm0.05$ \\
WJ1343167+394508 & L5 & K00 & $29\pm4^2$ &17.47$\pm0.35^4$  &3.98 & 13.49 & $2.07\pm0.10$ \\
WJ1439284+192915 & L1 & K99 & 69.8$\pm0.6^1$ & 15.34$\pm0.18^3$ & 3.43 & 11.91$^3$ & 1.20$\pm0.03$ \\
WJ1507476-162738 & L5 & R00 &131.0$\pm23.2^1$ & 17.24$\pm0.35^3$ & 3.83 & 13.41$^3$& 1.42$\pm0.03$ \\
WJ1632291+190441 & L8 & K99 & $59.5\pm2.9^1$ & $18.86\pm0.11^3$ & 4.18 & 14.68$^3$ & $1.86\pm0.05$ \\
WJ1726000+153819 & L2 & K00 & 50$\pm7.5^2$  & 18.05$\pm0.35^4$ & 3.91  & 14.14$^4$& 2.01$\pm0.08$ \\ 
\tableline\tableline
\end{tabular}
\end{center}
References: K99 - Kirkpatrick {\sl et al.}, 1999a; \\
K00 - Kirkpatrick {\sl et al.}, 2000; R00 - Reid {\sl et al.} (2000). \\
Notes: $^1$ USNO trigonometric  parallax (Dahn {\sl et al.}, 1999); \\
$^2$ Photometric parallax estimate (K99/K00); \\
$^3$ I$_C$, J and K photometry from USNO observations (Dahn {\sl et al.}, 1999); \\
$^4$ I$_C$ computed from I$_{814}$ using the relation given in section 2; \\
$^5$ I$_C$ from Dahn, priv. comm. 
\end{table}

\clearpage

\begin{table}
\begin{center}
{\bf Table 2:  Journal of observations}
\begin{tabular}{lcccccccc}
\tableline\tableline
Source & Epoch & F606W & F814W & l &b  & N$_b^1$\\
 & & Exposure & Exposure &  \\
\tableline
2M0036 & 15/02/2000 & 100s & $3 \times 100$s &119 & -44$^o$ & 2\\
2M0708 & 23/03/2000 & 100s &  300s, 350s& 188 & 17 & 10 \\
2M0740 & 27/03/2000 & 110s & 300s, 350s & 188 & 25 & 14\\
2M0746 & 15/04/2000 & 50s & $3 \times 50$s & 201 & 21 & 5\\
2M0820 & 24/04/2000 & 100s & 300s, 350s & 175 & 35 & 7\\
2M0825 & 25/03/2000 & 100s & 300s, 350s & 203 & 30 & 10\\
2M0850 & 1/02/2000 & 100s & 300s, 350s&209  & 32 & 3\\
2M0913 & 5/04/2000 & 100s & 300s, 350s & 211 & 40 & 5\\
2M0920 & 9/02/2000 & 100s & 300s, 350s&189  & 45 & 16\\
2M0928 & 28/04/2000 & 100s & 300s, 350s & 249 & 25 & 3\\
2M1123 & 19/04/2000 & 100s & 300s, 350s & 169 & 68 & 5\\
2M1146 & 28/04/2000 & 100s & 300s, 350s & 228 & 75 & 6\\
2M1155 & 18/03/2000 & 100s & 300s, 350s & 229 & 77 & 8\\
2M1328 & 23/04/2000 & 100s & 300s, 350s & 0 & 79 & 4\\
2M1338 & 25/04/2000 & 100s & 300s, 350s & 91 & 73 & 3\\
2M1343 & 21/04/2000 & 100s & 300s, 350s & 84 & 73 & 5\\
2M1439 & 22/03/2000 & 100s & $3 \times 100s$ & 21 & 63 & 5\\
2M1507 & 24/02/2000 & 100s & $3 \times 140$s &89 &  34 & 7 \\
2M1632 & 20/04/2000 & 100s & 300s, 350s & 36 & 39 & 6\\
2M1726 & 24/03/2000 & 100s &  300s, 350s & 28 & 25 & 14\\
\tableline\tableline
\end{tabular}
\end{center}
$^1$ N$_b$ is the number of stellar sources in each PC frame with I$_{814} > 20.5$ mag
\end{table}

\clearpage

\begin{table}
\begin{center}
{\bf Table 3: Binary parameters}
\begin{tabular}{lccccccccc}
\tableline\tableline
Source & $\Delta$  & PA (deg.) &M$_I$ & BC$_I$& M$_{bol}$ & M$_J$ 
& M$_{bol}$ & $\langle M_{bol} \rangle$ &  log($L \over L_\odot$) \\
\tableline
2M0746 & \\
AB & 0".22 & 15 & 14.67 & & & 11.31 \\
A &2.7 AU& & 15.17 &1.5 & 13.7 & 11.85 & 13.75 & 13.75 &-3.6$\pm0.1$ \\
B && & 15.79 & 1.8 & 14.0 & 12.32 & 14.2 & 14.1 & -3.75$\pm0.1$\\
2M0850 & \\
AB & 0".16  &250 & 18.21 &  & & 13.90 & \\ 
A &4.4 AU& & 18.50  & 2.4 & 16.1 & 14.3 & 16.2 & 16.15 & -4.6$\pm0.1$ \\
B && & 19.84 & 3.0 & 16.8 & 15.2 & 17.1 & 16.95& $-4.9\pm0.1$ \\
2M0920 & \\
AB & 0".07 & 90 & 17.68&  &  & 13.95  \\
A &1.6 AU& & 18.26  & 2.3 & 15.95 & 14.55 & 16.45 & 16.2 & -4.6$\pm0.1$ \\
B && & 18.70  & 2.6 & 16.1 & 14.9 & 16.8 & 16.35 & -4.65$\pm0.1$ \\
2M1146 & \\
AB & 0".29 &199  & 15.76 & & & 12.03 \\
A &7.6 AU& & 16.37 & 1.8 & 14.6 & 12.67 & 14.6 & 14.6 & -3.95$\pm0.1$ \\
B && & 16.68 & 2.0 & 14.7 & 12.90 & 14.8 & 14.75 & -4.0$\pm0.1$ \\
\tableline\tableline
\end{tabular}
\end{center}
\end{table}

\clearpage

\begin{table}
\begin{center}
{\bf Table 4: Mass estimates \\}
\begin{tabular}{lccccc}
\tableline\tableline
2M0850 & \\
Age (Gyrs) & M$_A (M_\odot)$ & M$_B (M_\odot)$ & T$_{eff}(A)$ & T$_{eff}(B)$ \\
0.19 & 0.02 & 0.015 & 1210 & 1140 \\
0.40 & 0.03 & 0.022 & 1260 & 1100 \\
0.71 & 0.04 & 0.031 & 1310 & 1140 \\
1.17 & 0.05 & 0.04 & 1350 & 1160 \\
1.72 & 0.06 & 0.047 & 1380 & 1230 \\
\tableline\tableline
\end{tabular}
\end{center}
\end{table}
\clearpage

\begin{table}
\begin{center}
{\bf Table 5: Known L dwarf/brown dwarf binaries\\}
\begin{tabular}{lccccl}
\tableline\tableline
System & M$_{pri}$ ($M_\odot$)& M$_{sec}$ ($M_\odot$) & $q^1$ & $\Delta$ AU & reference \\
\tableline 
PPl 15$^2$ &  0.07 & 0.06 & 0.86 & 0.03 & Basri \& Mart\'in, 1999 \\
HD 10697 & 1.10 & 0.04 & 0.035 & 0.07 & Shay \& Mazeh, 2000 \\
2M0746 & $>$0.06 & $>0.06$ & 1.0 & 2.7 & this paper \\
2M0920 & 0.06-0.075 & 0.06-0.075 & 0.95 & 3.2 & this paper \\
2M0850 & $<0.06$ & $<0.06$ & 0.75 & 4.4 & this paper \\
DENIS 1228&  $<0.06$ & $<0.06$ & $\sim 1$ & 4.9 & Mart\'in {\sl et al.}, 1999 \\
2M1146 & $<0.06$ & $<0.06$ & $\sim 1$ & 7.6 & Ko99 \\
DENIS 0205 & 0.06-0.09 & 0.06-0.09 &  $\sim 1$ & 9.2 & Ko99 \\
Gl 229B & 0.5 & $\sim0.045$ & $\sim 0.1$ & 44 & Nakajima {\sl et al.}, 1995 \\
TWA 5$^{2,3}$ & 0.4 & 0.025 & 0.06 & 100 & Lowrance {\sl et al.}, 1999\\
GD 165B & $> 1$ & $<0.08$ & $<.08$& 110 & Becklin \& Zuckerman, 1988 \\  
HR 7329B & $\sim5$ & $<0.05$ & $<.01$ & 200 & Lowrance {\sl et al.}, 2000 \\
GJ 1048B & $\sim0.7$ & $<0.08$ & $<0.11$ & 250 & Gizis {\sl et al.}, 2000 \\
G196-3B & 0.5 & $\sim0.025$ & $\sim 0.05$ & 340 & Rebolo {\sl et al.}, 1998 \\
GJ 1001B & 0.4 & $\sim0.05$ & $\sim 0.13$ & 180 & Goldman {\sl et al.}, 1999\\
Gl 570D & 0.7 & $\sim0.05$ & $\sim 0.07$ & 1525 & Burgasser {\sl et al.}, 2000b \\
Gl 417B & 1.0  & $\sim0.035$& $\sim 0.035$ & 2000 & Kirkpatrick {\sl et al.}, in prep. \\
Gl 584C & 1.0 & $\sim0.060$& $\sim 0.060$ & 3600 & Kirkpatrick {\sl et al.}, in prep. \\
\tableline\tableline
\end{tabular}
\end{center}
$^1$ Mass ratios for L dwarf/L dwarf systems are based on the relative K-band luminosity

$^2$ Members of Pleiades cluster or TW Hydrae association

$^3$ High resolution spectroscopy indicates that several other stars in this
moving group are binary or multiple systems.
\end{table}
\clearpage

\begin{figure}
\plotone{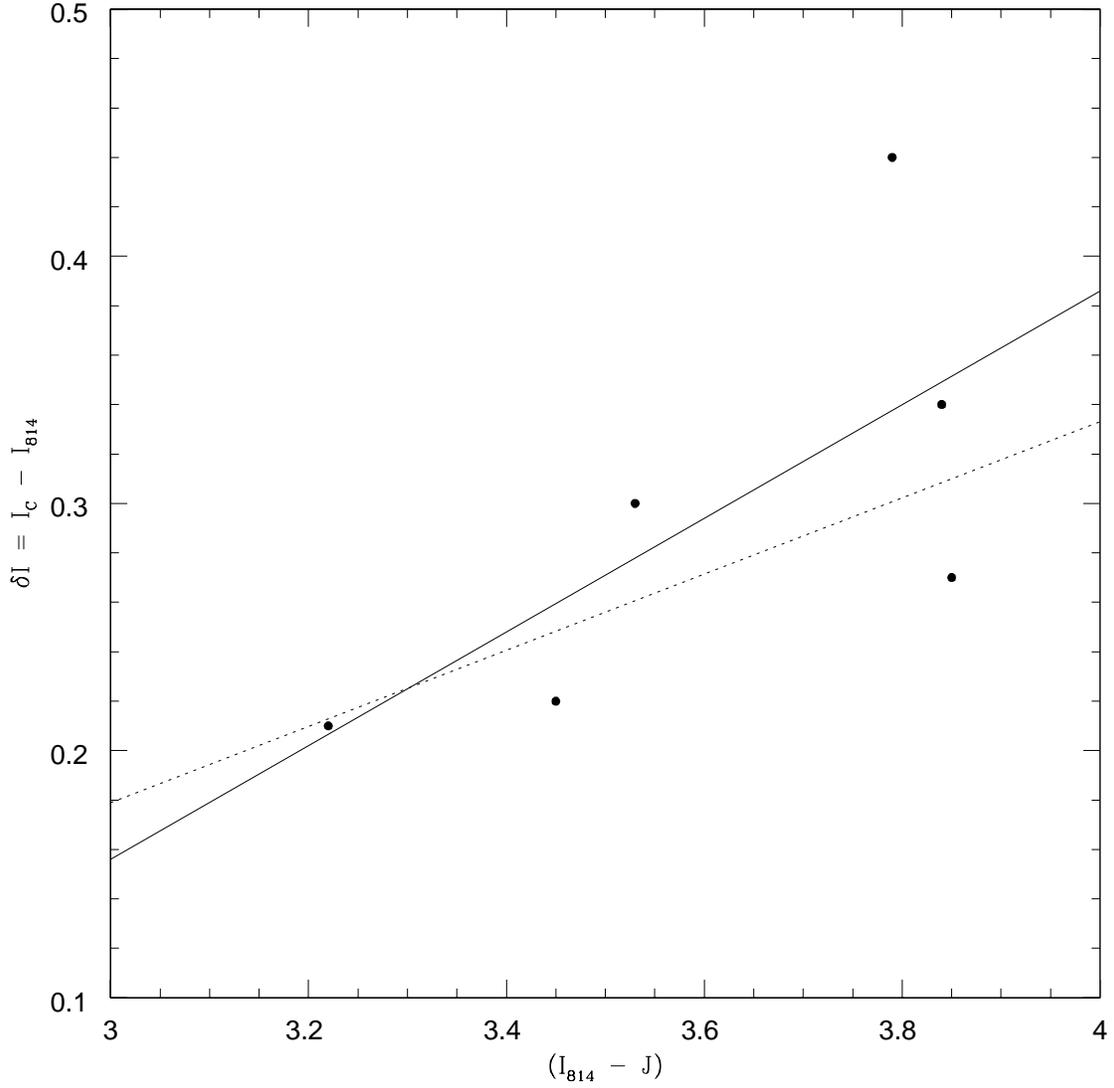}
\caption{ The offset between I$_C$ and I$_{814}$ as a function of (I$_{814}$-J) colour
for L dwarfs. The solid line plots the best-fit linear relation, the dashed line plots
the linear regression if the most discrepant point is omitted from the sample..}
\end{figure}

\begin{figure}
\plotone{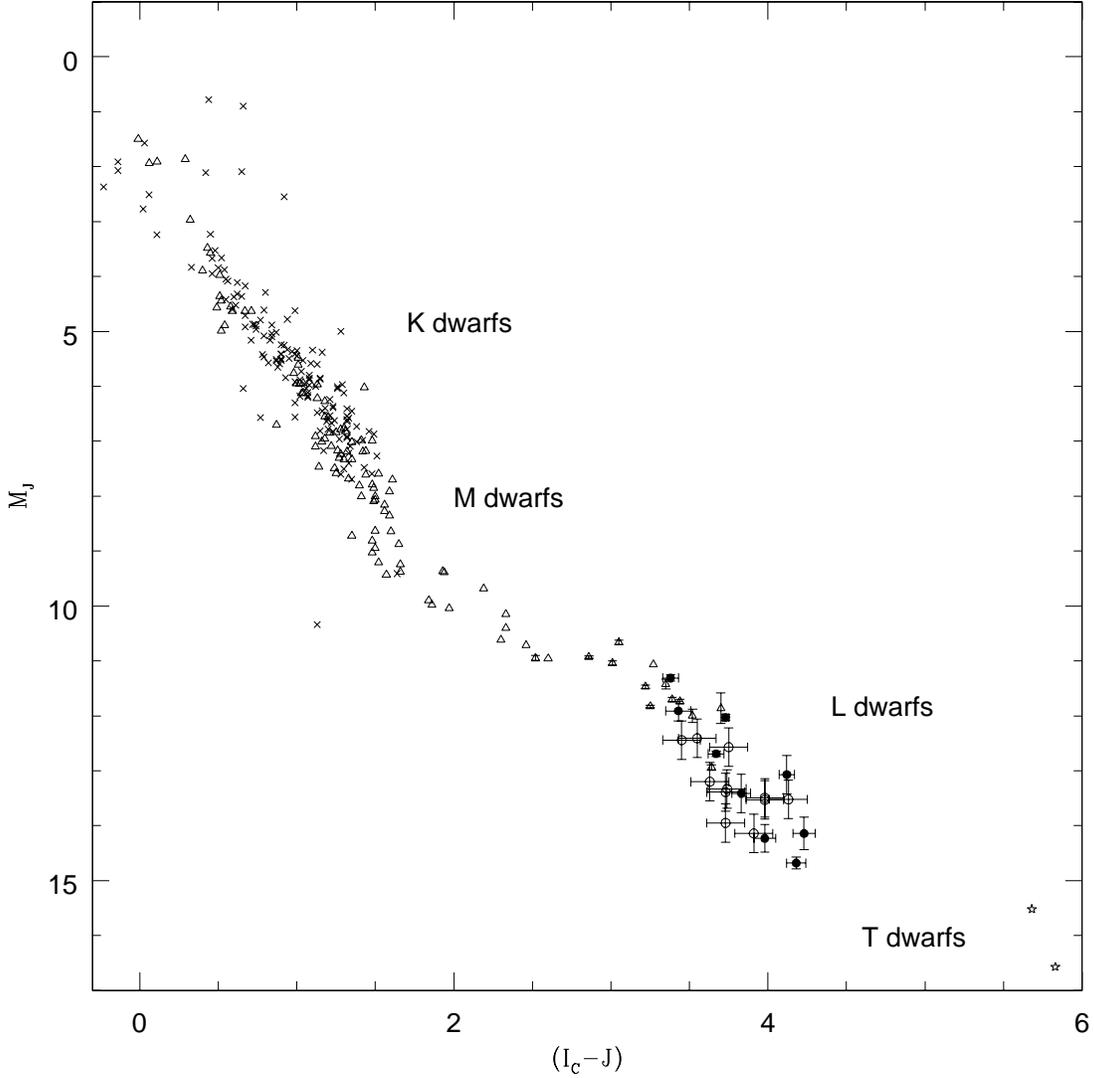}
\caption{The (M$_J$, (I$_C$-J)) diagram for low-mass dwarfs. Crosses are nearby stars with BVRI
data from Bessell (1989), JHK observations from 2MASS and parallax measurements from Hipparcos; 
open triangles mark nearby, single  stars with accurate trigonometric parallax 
measurements (from Reid \& Gizis, 1997b), and late-M/L dwarfs from the USNO parallax program (Dahn
{\sl et al.}, in prep); solid points identify the nine L dwarfs in the present HST sample which have trigonometric
parallax measurements (Dahn {\sl et al.});  the remaining 11  HST L dwarfs, with spectroscopic
parallax estimates, are plotted as open circles; finally. five-pointed stars mark data
for the T dwarfs Gl 229B (Nakajima {\sl et al.}, 1995) and Gl 570D (Burgasser {\sl et al.}, 2000).}
\end{figure}

\begin{figure}
\caption{A: HST Planetary Camera images of 2M0746. The left
panel shows the F606W exposure; the right panel the F814W image. The
field of view is $6 \times 6$ arcseconds. (Figures enclosed separately.)}
\end{figure}   

\setcounter{figure}{2}
\begin{figure}
\caption{B: HST Planetary Camera images of 2M0850. The left
panel shows the F606W exposure; the right panel the F814W exposure. The 
bright object in the upper left corner is an unrelated M dwarf.}
\end{figure}   

\setcounter{figure}{2}
\begin{figure}
\caption{C: HST Planetary Camera images of 2M0920. The left
panel shows the F606W exposure; the right panel the F814W exposure.}
\end{figure}   

\setcounter{figure}{2}
\begin{figure}
\caption{D: HST Planetary Camera images of 2M1146. The left
panel shows the F606W exposure; the right panel the F814W exposure.}
\end{figure}   

\clearpage

\begin{figure}
\plotone{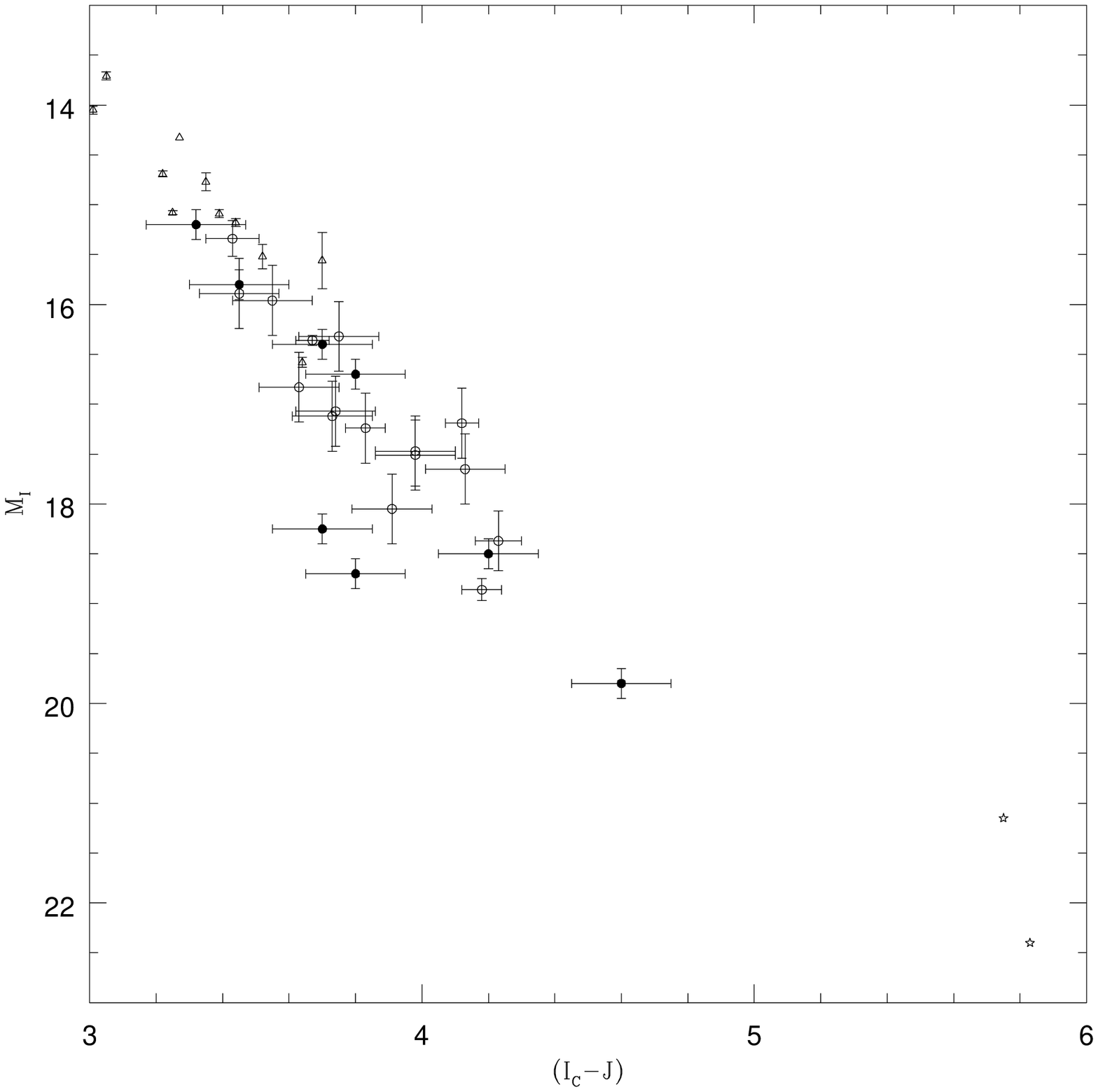}
\caption{ The (M$_I$, (I$_C$-J)) colour-magnitude diagram for ultracool dwarfs. As in figure
2, open triangles mark data for nearby late-type M dwarfs and L dwarfs with
trigonometric parallax measurements. The open circles plot data for the apparently
single L dwarfs with HST observations; solid points (with errorbars) mark our
estimates of the location of the four binary components. }
\end{figure}
\clearpage

\begin{figure}
\plotone{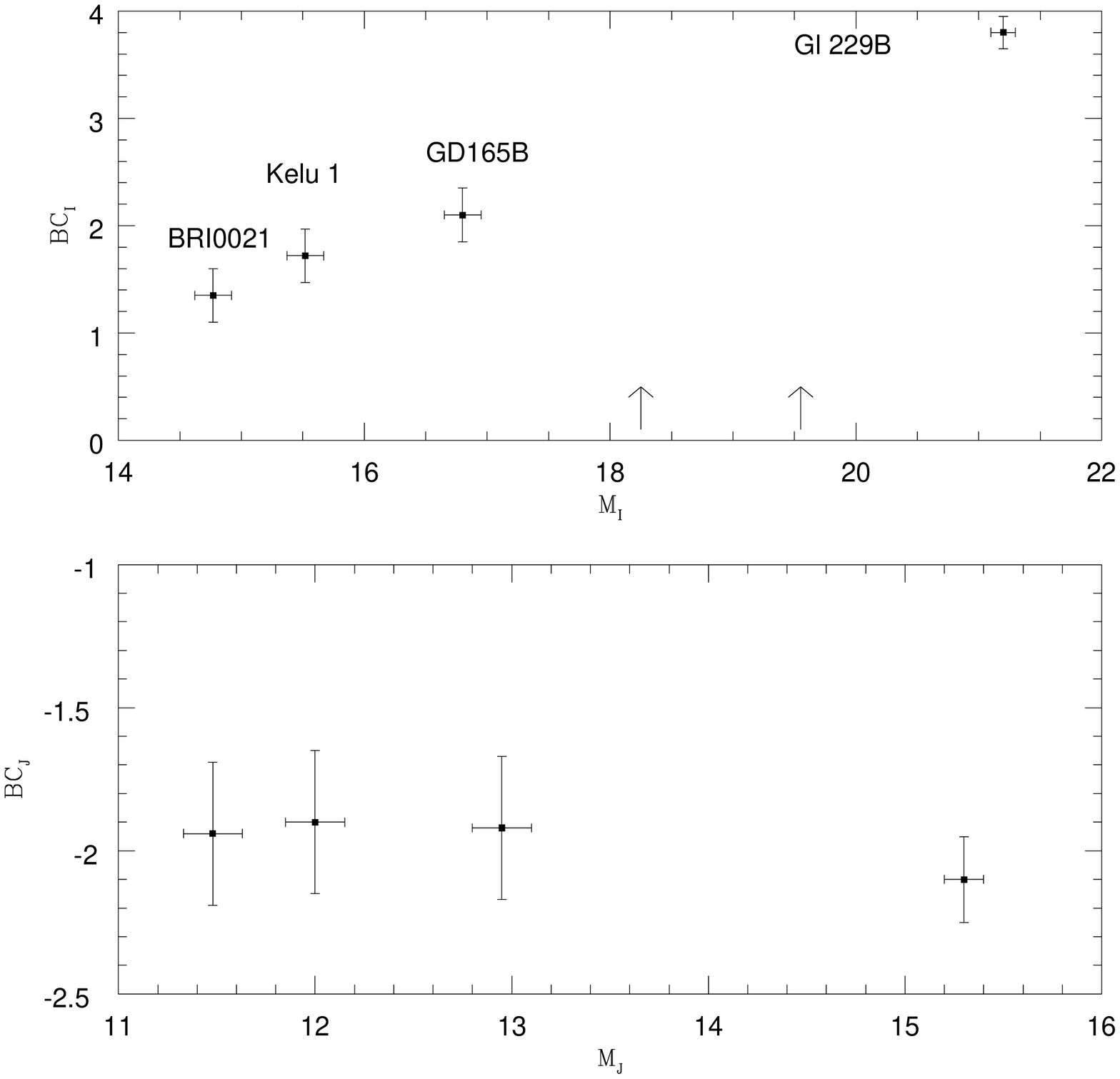}
\caption{ Bolometric corrections for late-type dwarfs. The upper
panel plots the available data for the Cousins I-band, where 
the arrows mark the locations of 2M0850A and 2M0850B; the 
lower panel plots (M$_J$, BC$_J$) data for the same four calibrators. }
\end{figure}
\clearpage

\begin{figure}
\plotone{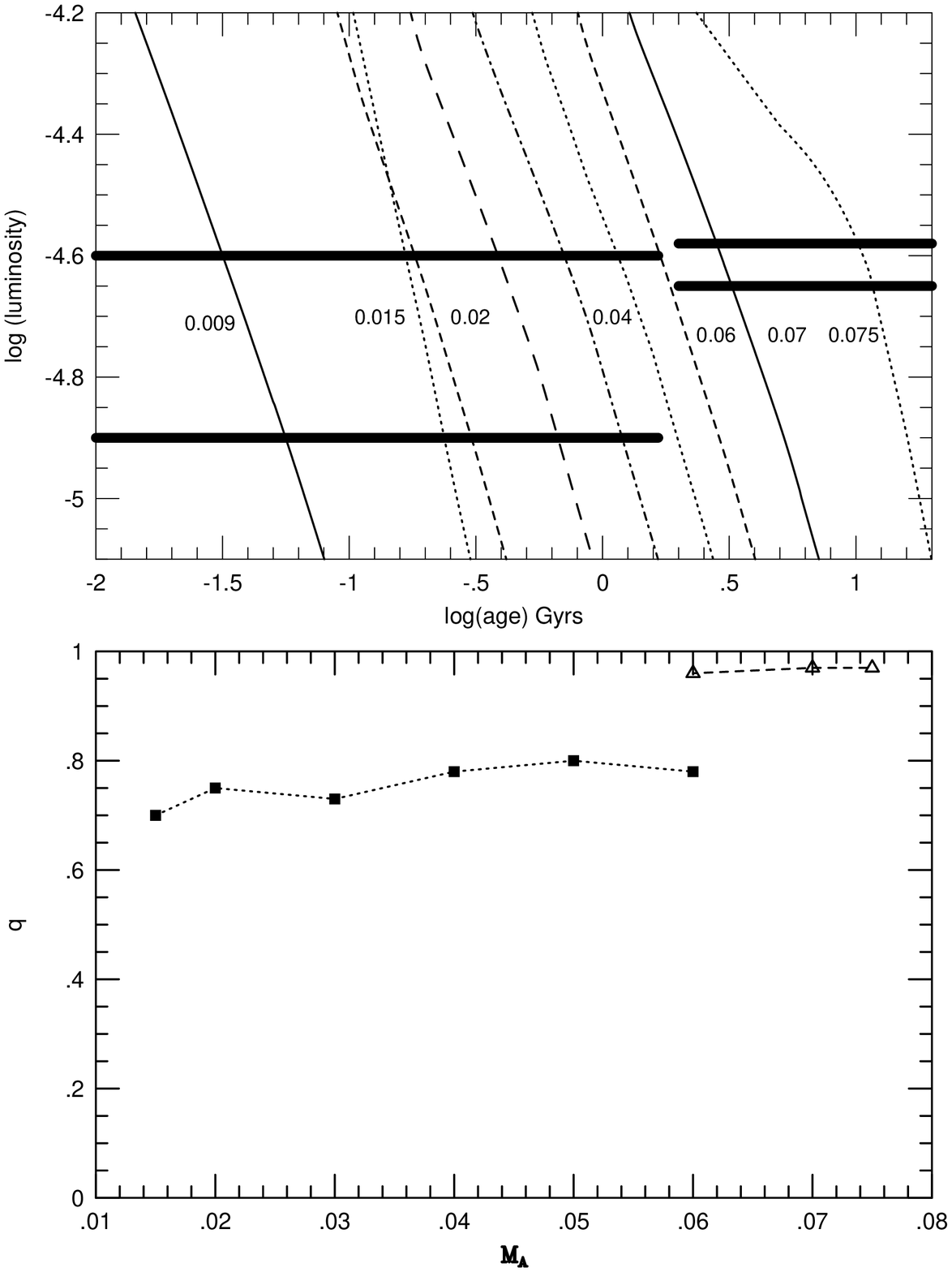}
\caption{Mass estimates for 2M0850AB and 2M0920AB. The upper
panel plots evolutionary tracks for brown dwarfs calculated by
Burrows {\sl et al.} (1993, 1997). The solid horizontal lines plot
the luminosity estimates for the four components: the left pair of
lines plot 2M0850A and B, with the upper mass limit for 2M0850A set at 0.06M$_\odot$
through the detection of lithium; the right-hand pair of lines, closer spaced in
luminosity, mark 2M0920A and B. In the latter case, the non-detection of
lithium sets a lower mass limit of 0.06M$_\odot$ for 2M0920B.
The lower panel plots the inferred mass ratios for the two systems, 
with squares marking 2M0850A/B and triangles for the near equal-mass 2M0920A/B system.}
\end{figure}   

\clearpage
\begin{figure}
\plotone{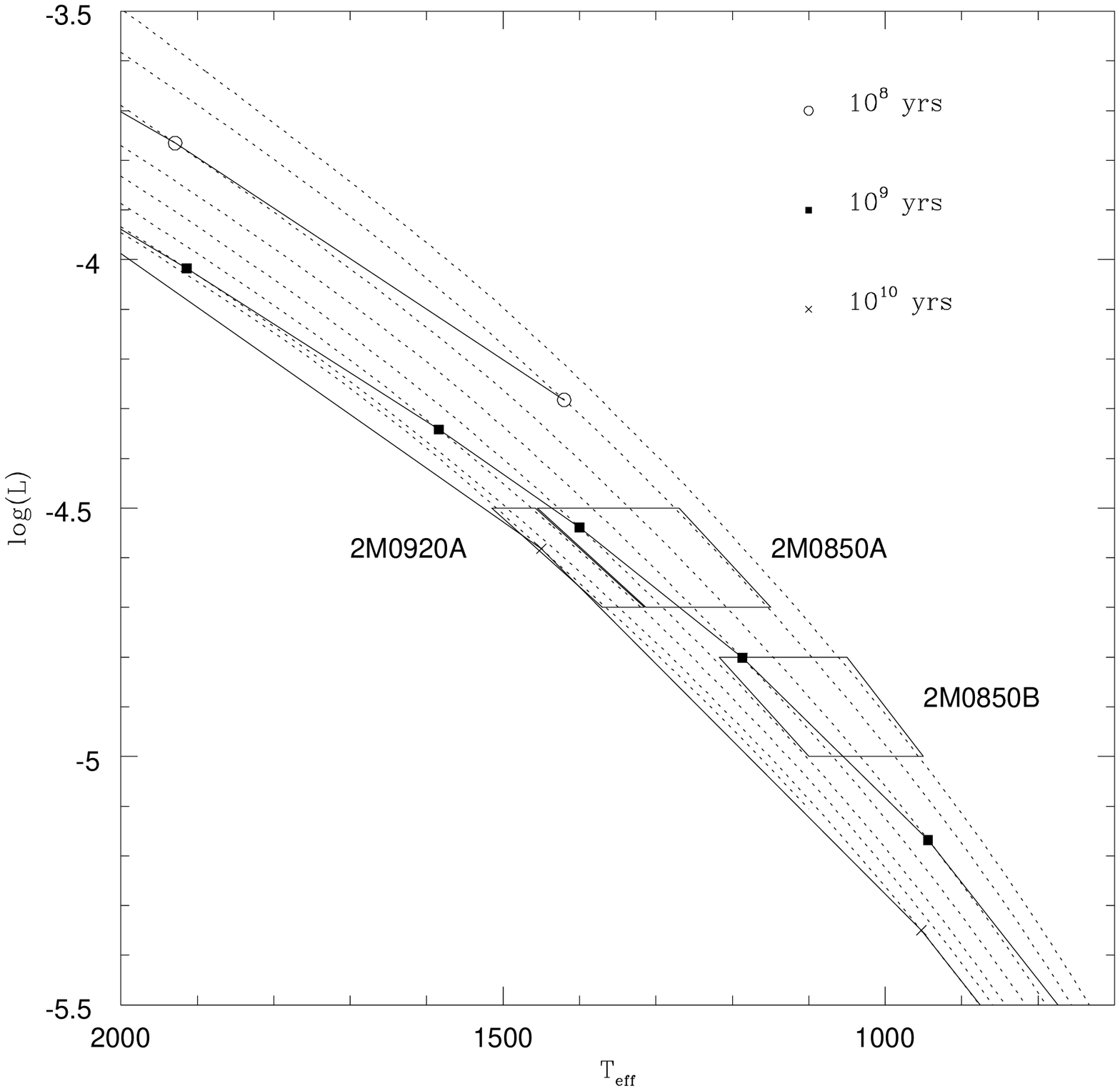}
\caption {Error boxes for the brown dwarf components of 2M0850AB and for
2M0920A (2M0920B is not plotted) superimposed on
evolutionary tracks for brown dwarfs from Burrows {\sl et al.} (1997). The
solid lines mark 10$^8$, 10$^9$ and 10$^{10}$ year isochrones; the 
dotted lines mark the evolution of 0.015, 0.02, 0.03, 0.04, 0.05, 0.06, 
0.07 and 0.075 M$_\odot$ models: the 0.015M$_\odot$ model has the highest 
luminosity at a given temperature.}
\end{figure}

\clearpage
\begin{figure}
\plotone{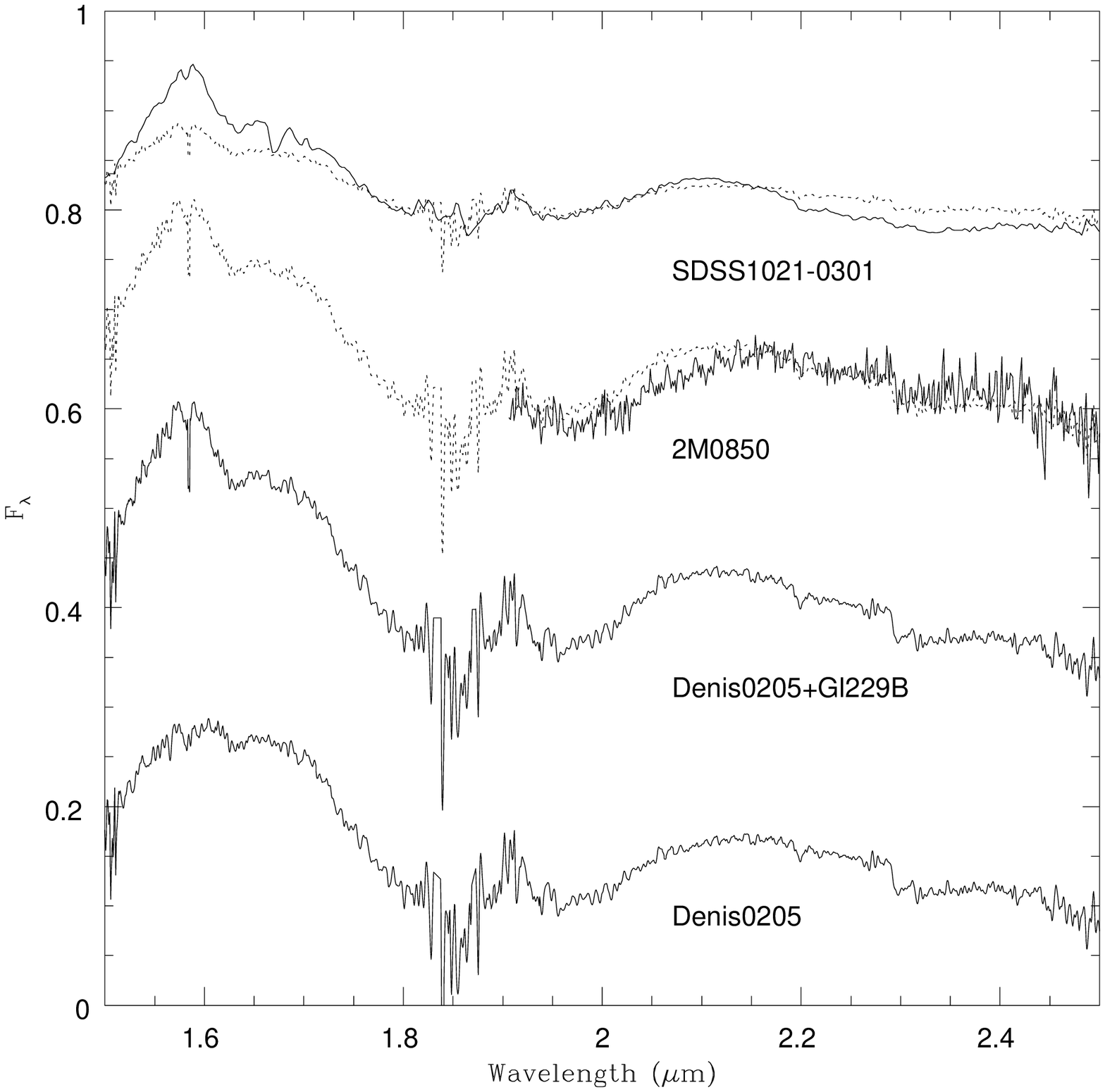}
\caption {The composite 1.5 to 2.5 $\mu$m  spectrum f a hypothetical late-L/T-dwarf binary. The lowest
spectrum plots our CGS4 data for the L7 dwarf, DENIS-P 0205.4-1159; next, we plot the hypothetical
binary, combining the DENIS0205 spectrum with data for Gl 229B (Geballe {\sl et al.}, 1996), 
with $\Delta J_{B-A} = 1$
magnitude; third from the bottom, we plot K-band data for 2M0850; the top spectrum plots CGS4 data for
the early T-dwarf, SDSS1021-0301 (data courtesy of S. Leggett). The dotted line superimposed on the
latter two spectra plots the L/T composite, scaled to match at 2.15 $\mu$m. }
\end{figure}

\clearpage
\begin{figure}
\plotone{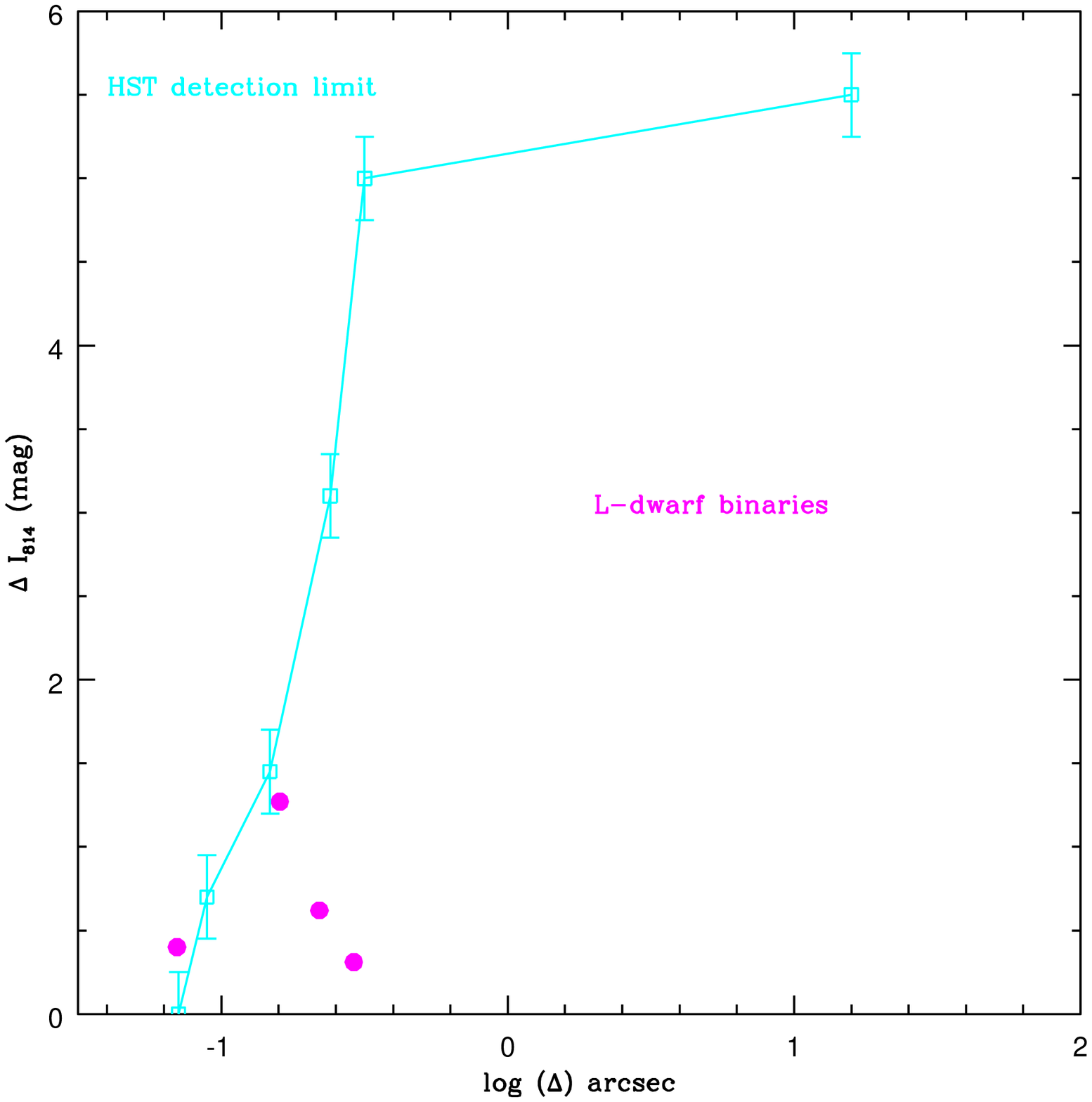}
\caption {A comparison between the observed primary-secondary magnitude difference, 
$\delta I_{F814}$, and the formal detection limits of the HST Planetary Camera
observations. }
\end{figure}

\clearpage
\begin{figure}
\plotone{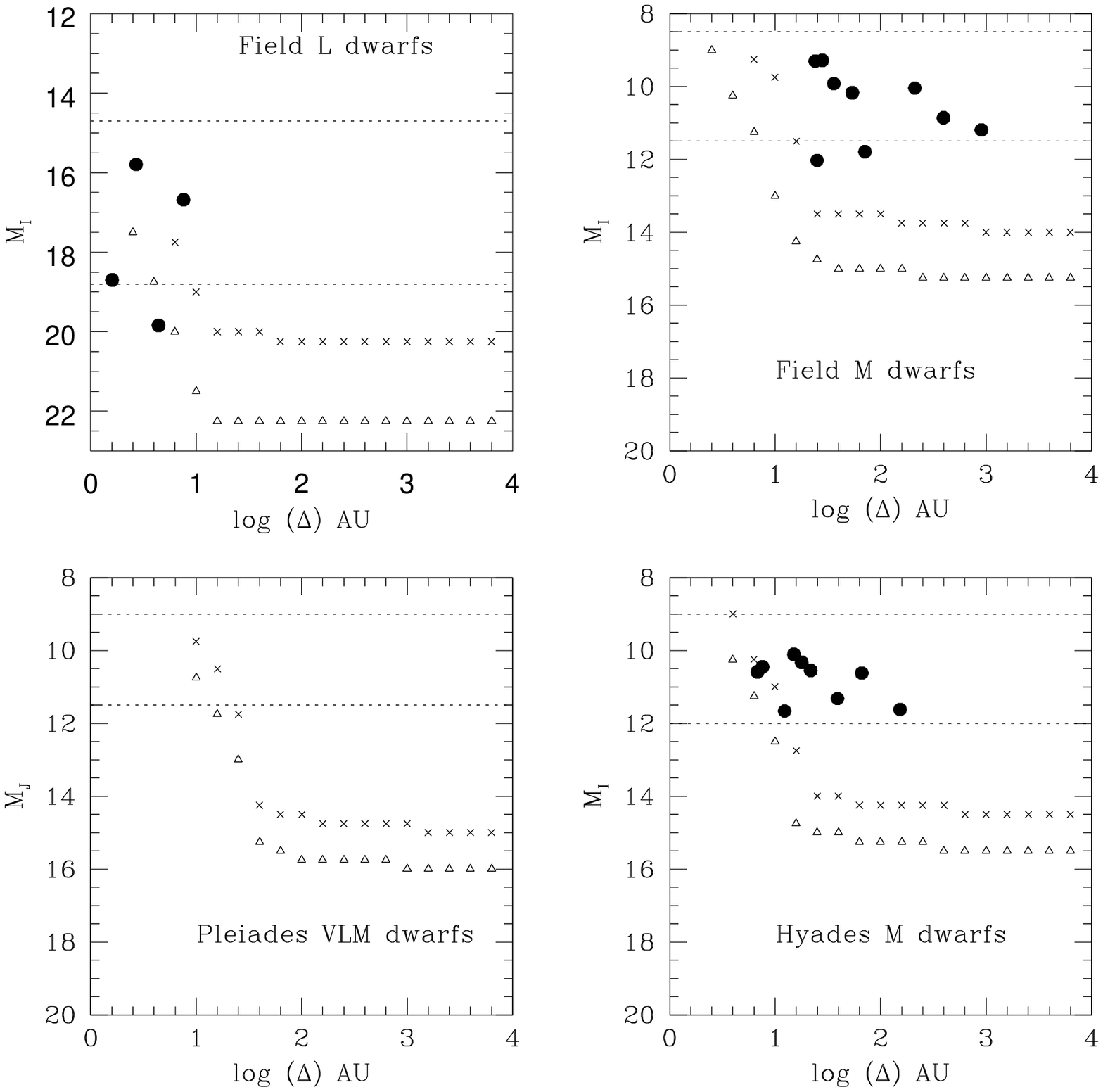}
\caption { Sensitivity limits of four high-resolution imaging surveys for companions. The L
dwarf data are from the present paper; field M dwarfs from Reid \& Gizis (1997a); 
Hyades M dwarfs from Reid \& Gizis (1997b); and Pleiades data from Mart\'in {\sl et al.} (2000).
Note that we plot M$_J$ for the last mentioned dataset.
Limiting magnitude (M$_I$ or M$_J$) is plotted as a function of linear separation from the primary (in AU): 
crosses mark the 100\% completeness level; open triangles, 50\% completeness. The horizontal
dotted lines indicate the approximate range of absolute magnitudes of the primaries, and detected
companions are plotted as solid points. }
\end{figure}

\clearpage
\begin{figure}
\plotone{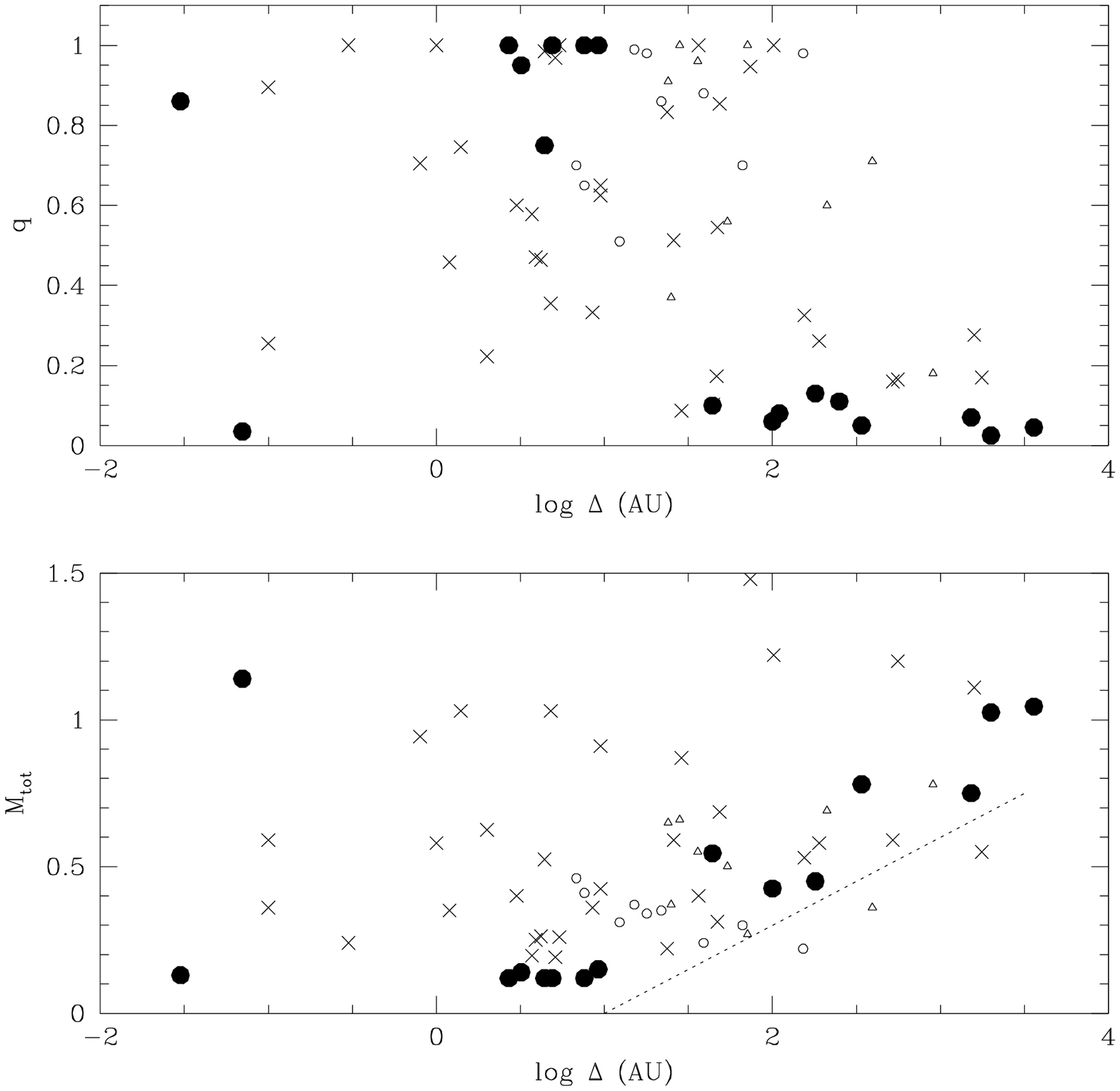}
\caption {The upper panel plots the (mass ratio, log(separation)) distribution for L dwarf binaries 
from Table 5 (solid point) matched against similar data for M dwarf binaries in the Solar Neighbourhood
(crosses) and from the HST Hyades (open circles) and field (open triangles) surveys. The lower
panel plots the total mass of each binary against log($\Delta$), using the same symbols.}
\end{figure}

\end{document}